\pdfoutput=1

\let\singlecol\undefined 

\ifdefined\singlecol
  \documentclass[useAMS,usenatbib,referee,longauth]{aa}
\else
  \documentclass[useAMS,usenatbib,longauth]{aa}
\fi

\usepackage[pdftex]{}
\usepackage{times}
\usepackage{wasysym}
\usepackage{url}
\usepackage[switch]{lineno}
\usepackage{amssymb}
\usepackage{subfigure}
\usepackage{amsmath}
\usepackage{siunitx}
\usepackage{xcolor}
\usepackage{float}

\usepackage{hyperref}
\hypersetup{
    colorlinks=true,
    linkcolor=blue,
    citecolor=blue,
    filecolor=blue,      
    urlcolor=blue,
}

\newcommand{\etacar}{$\eta$~Car}
\newcommand{\etacare}{$\eta$~Carinae~}

\newcommand{\UNITS}[1]{\,\mathrm{#1}}

\newcommand{\grad}{$^{\circ}$}

\newcommand{\fermi}{\emph{Fermi}-LAT}

\newcommand{\hess}{H.E.S.S.}

\addto\extrasenglish{%

}

\DeclareSIUnit\solarmass{\ensuremath{M_\odot}}
\DeclareSIUnit\solarradius{\ensuremath{R_\odot}}
\DeclareSIUnit\year{yr}
\DeclareSIUnit\solarluminosity{\ensuremath{L_\odot}}
\DeclareSIUnit\gauss{G}
\DeclareSIUnit\erg{erg}
\DeclareSIUnit\pc{pc}
\def\@fnsymbol#1{\ensuremath{\ifcase#1\or *\or \dagger\or \ddagger\or
   \mathsection\or \mathparagraph\or \|\or **\or \dagger\dagger
   \or \ddagger\ddagger \else\@ctrerr\fi}}

\begin{document} 

\title{Detection of very-high-energy gamma-ray emission from Eta Carinae during its 2020 periastron passage}
\date{Received 18 November 2024 / Accepted 20 January 2025}
\author{H.E.S.S. Collaboration
\and F.~Aharonian \inst{\ref{DIAS},\ref{MPIK},\ref{Yerevan}}
\and F.~Ait~Benkhali \inst{\ref{LSW}}
\and J.~Aschersleben \inst{\ref{Groningen}}
\and H.~Ashkar \inst{\ref{LLR}}
\and V.~Barbosa~Martins \inst{\ref{DESY}}
\and R.~Batzofin \inst{\ref{UP}}
\and Y.~Becherini \inst{\ref{APC},\ref{Linnaeus}}
\and D.~Berge \inst{\ref{DESY},\ref{HUB}}
\and K.~Bernl\"ohr \inst{\ref{MPIK}}
\and M.~B\"ottcher \inst{\ref{NWU}}
\and C.~Boisson \inst{\ref{LUTH}}
\and J.~Bolmont \inst{\ref{LPNHE}}
\and M.~de~Bony~de~Lavergne \inst{\ref{CEA}}
\and F.~Bradascio \inst{\ref{CEA}}
\and R.~Brose \inst{\ref{DIAS}}
\and A.~Brown \inst{\ref{Oxford}}
\and F.~Brun \inst{\ref{CEA}}
\and B.~Bruno \inst{\ref{ECAP}}
\and C.~Burger-Scheidlin \inst{\ref{DIAS}}
\and S.~Casanova \inst{\ref{IFJPAN}}
\and J.~Celic \inst{\ref{ECAP}}
\and M.~Cerruti \inst{\ref{APC}}
\and T.~Chand \inst{\ref{NWU}}
\and S.~Chandra \inst{\ref{NWU}}
\and A.~Chen \inst{\ref{Wits}}
\and J.~Chibueze \inst{\ref{NWU}}
\and O.~Chibueze \inst{\ref{NWU}}
\and T.~Collins \inst{\ref{UP}}
\and G.~Cotter \inst{\ref{Oxford}}
\and J.~Damascene~Mbarubucyeye \inst{\ref{DESY}}
\and J.~de~Assis~Scarpin \inst{\ref{LLR}}
\and J.~Devin \inst{\ref{LUPM}}
\and A.~Djannati-Ata\"i \inst{\ref{APC}}
\and J.~Djuvsland \inst{\ref{MPIK}}
\and A.~Dmytriiev \inst{\ref{NWU}}
\and K.~Egberts \inst{\ref{UP}}
\and S.~Einecke \inst{\ref{Adelaide}}
\and J.-P.~Ernenwein \inst{\ref{CPPM}}
\and C.~Esca\~{n}uela~Nieves \inst{\ref{MPIK}}
\and K.~Feijen \inst{\ref{APC}}
\and M.~Filipovic \inst{\ref{Sydney}}
\and G.~Fontaine \inst{\ref{LLR}}
\and S.~Funk \inst{\ref{ECAP}}
\and S.~Gabici \inst{\ref{APC}}
\and J.F.~Glicenstein \inst{\ref{CEA}}
\and G.~Grolleron \inst{\ref{LPNHE}}
\and M.-H.~Grondin \inst{\ref{CENBG}}
\and L.~Haerer \inst{\ref{MPIK}}
\and B.~He\ss \inst{\ref{IAAT}}
\and J.A.~Hinton\protect\footnotemark[1] \inst{\ref{MPIK}}
\and W.~Hofmann \inst{\ref{MPIK}}
\and T.~L.~Holch \inst{\ref{DESY}}
\and M.~Holler \inst{\ref{Innsbruck}}
\and D.~Horns \inst{\ref{UHH}}
\and Zhiqiu~Huang \inst{\ref{MPIK}}
\and M.~Jamrozy \inst{\ref{UJK}}
\and F.~Jankowsky \inst{\ref{LSW}}
\and A.~Jardin-Blicq \inst{\ref{CENBG}}
\and I.~Jung-Richardt \inst{\ref{ECAP}}
\and K.~Katarzy{\'n}ski \inst{\ref{NCUT}}
\and R.~Khatoon \inst{\ref{NWU}}
\and B.~Kh\'elifi \inst{\ref{APC}}
\and W.~Klu\'{z}niak \inst{\ref{NCAC}}
\and Nu.~Komin \inst{\ref{Wits}}
\and K.~Kosack \inst{\ref{CEA}}
\and D.~Kostunin \inst{\ref{DESY}}
\and R.G.~Lang \inst{\ref{ECAP}}
\and S.~Le~Stum \inst{\ref{CPPM}}
\and A.~Lemi\`ere \inst{\ref{APC}}
\and M.~Lemoine-Goumard \inst{\ref{CENBG}}
\and J.-P.~Lenain \inst{\ref{LPNHE}}
\and A.~Luashvili \inst{\ref{LUTH}}
\and J.~Mackey \inst{\ref{DIAS}}
\and D.~Malyshev \inst{\ref{IAAT}}
\and V.~Marandon\protect\footnotemark[1] \inst{\ref{CEA}}
\and A.~Marcowith \inst{\ref{LUPM}}
\and G.~Mart\'i-Devesa \inst{\ref{Innsbruck}}
\and R.~Marx \inst{\ref{LSW}}
\and A.~Mehta \inst{\ref{DESY}}
\and A.~Mitchell \inst{\ref{ECAP}}
\and R.~Moderski \inst{\ref{NCAC}}
\and M.O.~Moghadam \inst{\ref{UP}}
\and L.~Mohrmann \inst{\ref{MPIK}}
\and E.~Moulin \inst{\ref{CEA}}
\and M.~de~Naurois \inst{\ref{LLR}}
\and J.~Niemiec \inst{\ref{IFJPAN}}
\and S.~Ohm \inst{\ref{DESY}}
\and L.~Olivera-Nieto \inst{\ref{MPIK}}
\and E.~de~Ona~Wilhelmi \inst{\ref{DESY}}
\and M.~Ostrowski \inst{\ref{UJK}}
\and S.~Panny \inst{\ref{Innsbruck}}
\and M.~Panter \inst{\ref{MPIK}}
\and R.D.~Parsons \inst{\ref{HUB}}
\and U.~Pensec \inst{\ref{LPNHE}}
\and G.~P\"uhlhofer \inst{\ref{IAAT}}
\and A.~Quirrenbach \inst{\ref{LSW}}
\and S.~Ravikularaman \inst{\ref{APC},\ref{MPIK}}
\and M.~Regeard \inst{\ref{APC}}
\and A.~Reimer \inst{\ref{Innsbruck}}
\and O.~Reimer \inst{\ref{Innsbruck}}
\and Q.~Remy\protect\footnotemark[1] \inst{\ref{MPIK}}
\and H.~Ren \inst{\ref{MPIK}}
\and B.~Reville\protect\footnotemark[1] \inst{\ref{MPIK}}
\and F.~Rieger \inst{\ref{MPIK}}
\and G.~Rowell \inst{\ref{Adelaide}}
\and B.~Rudak \inst{\ref{NCAC}}
\and E.~Ruiz-Velasco \inst{\ref{MPIK}}
\and K.~Sabri \inst{\ref{LUPM}}
\and V.~Sahakian \inst{\ref{Yerevan}}
\and H.~Salzmann \inst{\ref{IAAT}}
\and A.~Santangelo \inst{\ref{IAAT}}
\and M.~Sasaki \inst{\ref{ECAP}}
\and J.~Sch\"afer \inst{\ref{ECAP}}
\and F.~Sch\"ussler \inst{\ref{CEA}}
\and H.M.~Schutte \inst{\ref{NWU}}
\and J.N.S.~Shapopi \inst{\ref{UNAM}}
\and S.~Spencer \inst{\ref{ECAP}}
\and {\L.}~Stawarz \inst{\ref{UJK}}
\and R.~Steenkamp \inst{\ref{UNAM}}
\and S.~Steinmassl\protect\footnotemark[1] \inst{\ref{MPIK}}
\and C.~Steppa \inst{\ref{UP}}
\and K.~Streil \inst{\ref{ECAP}}
\and T.~Tanaka \inst{\ref{Konan}}
\and R.~Terrier \inst{\ref{APC}}
\and M.~Tluczykont \inst{\ref{UHH}}
\and M.~Tsirou \inst{\ref{DESY}}
\and N.~Tsuji \inst{\ref{RIKKEN}}
\and C.~van~Eldik \inst{\ref{ECAP}}
\and M.~Vecchi \inst{\ref{Groningen}}
\and C.~Venter \inst{\ref{NWU}}
\and T.~Wach \inst{\ref{ECAP}}
\and S.J.~Wagner \inst{\ref{LSW}}
\and F.~Werner \inst{\ref{MPIK}}
\and R.~White \inst{\ref{MPIK}}
\and A.~Wierzcholska \inst{\ref{IFJPAN}}
\and M.~Zacharias \inst{\ref{LSW},\ref{NWU}}
\and A.A.~Zdziarski \inst{\ref{NCAC}}
\and A.~Zech \inst{\ref{LUTH}}
\and N.~\.Zywucka \inst{\ref{NWU}}
}

\institute{
Dublin Institute for Advanced Studies, 31 Fitzwilliam Place, Dublin 2, Ireland \label{DIAS} \and
Max-Planck-Institut f\"ur Kernphysik, Saupfercheckweg 1, 69117 Heidelberg, Germany \label{MPIK} \and
Yerevan State University,  1 Alek Manukyan St, Yerevan 0025, Armenia \label{Yerevan} \and
Landessternwarte, Universit\"at Heidelberg, K\"onigstuhl, D 69117 Heidelberg, Germany \label{LSW} \and
Kapteyn Astronomical Institute, University of Groningen, Landleven 12, 9747 AD Groningen, The Netherlands \label{Groningen} \and
Laboratoire Leprince-Ringuet, École Polytechnique, CNRS, Institut Polytechnique de Paris, F-91128 Palaiseau, France \label{LLR} \and
Deutsches Elektronen-Synchrotron DESY, Platanenallee 6, 15738 Zeuthen, Germany \label{DESY} \and
Institut f\"ur Physik und Astronomie, Universit\"at Potsdam,  Karl-Liebknecht-Strasse 24/25, D 14476 Potsdam, Germany \label{UP} \and
Université Paris Cité, CNRS, Astroparticule et Cosmologie, F-75013 Paris, France \label{APC} \and
Department of Physics and Electrical Engineering, Linnaeus University,  351 95 V\"axj\"o, Sweden \label{Linnaeus} \and
Institut f\"ur Physik, Humboldt-Universit\"at zu Berlin, Newtonstr. 15, D 12489 Berlin, Germany \label{HUB} \and
Centre for Space Research, North-West University, Potchefstroom 2520, South Africa \label{NWU} \and
Laboratoire Univers et Théories, Observatoire de Paris, Université PSL, CNRS, Université Paris Cité, 5 Pl. Jules Janssen, 92190 Meudon, France \label{LUTH} \and
Sorbonne Universit\'e, CNRS/IN2P3, Laboratoire de Physique Nucl\'eaire et de Hautes Energies, LPNHE, 4 place Jussieu, 75005 Paris, France \label{LPNHE} \and
IRFU, CEA, Universit\'e Paris-Saclay, F-91191 Gif-sur-Yvette, France \label{CEA} \and
University of Oxford, Department of Physics, Denys Wilkinson Building, Keble Road, Oxford OX1 3RH, UK \label{Oxford} \and
Friedrich-Alexander-Universit\"at Erlangen-N\"urnberg, Erlangen Centre for Astroparticle Physics, Nikolaus-Fiebiger-Str. 2, 91058 Erlangen, Germany \label{ECAP} \and
Instytut Fizyki J\c{a}drowej PAN, ul. Radzikowskiego 152, 31-342 Krak{\'o}w, Poland \label{IFJPAN} \and
School of Physics, University of the Witwatersrand, 1 Jan Smuts Avenue, Braamfontein, Johannesburg, 2050 South Africa \label{Wits} \and
Laboratoire Univers et Particules de Montpellier, Universit\'e Montpellier, CNRS/IN2P3,  CC 72, Place Eug\`ene Bataillon, F-34095 Montpellier Cedex 5, France \label{LUPM} \and
School of Physical Sciences, University of Adelaide, Adelaide 5005, Australia \label{Adelaide} \and
Aix Marseille Universit\'e, CNRS/IN2P3, CPPM, Marseille, France \label{CPPM} \and
School of Science, Western Sydney University, Locked Bag 1797, Penrith South DC, NSW 2751, Australia \label{Sydney} \and
Universit\'e Bordeaux, CNRS, LP2I Bordeaux, UMR 5797, F-33170 Gradignan, France \label{CENBG} \and
Institut f\"ur Astronomie und Astrophysik, Universit\"at T\"ubingen, Sand 1, D 72076 T\"ubingen, Germany \label{IAAT} \and
Universit\"at Innsbruck, Institut f\"ur Astro- und Teilchenphysik, Technikerstraße 25, 6020 Innsbruck, Austria \label{Innsbruck} \and
Universit\"at Hamburg, Institut f\"ur Experimentalphysik, Luruper Chaussee 149, D 22761 Hamburg, Germany \label{UHH} \and
Obserwatorium Astronomiczne, Uniwersytet Jagiello{\'n}ski, ul. Orla 171, 30-244 Krak{\'o}w, Poland \label{UJK} \and
Institute of Astronomy, Faculty of Physics, Astronomy and Informatics, Nicolaus Copernicus University,  Grudziadzka 5, 87-100 Torun, Poland \label{NCUT} \and
Nicolaus Copernicus Astronomical Center, Polish Academy of Sciences, ul. Bartycka 18, 00-716 Warsaw, Poland \label{NCAC} \and
University of Namibia, Department of Physics, Private Bag 13301, Windhoek 10005, Namibia \label{UNAM} \and
Department of Physics, Konan University, 8-9-1 Okamoto, Higashinada, Kobe, Hyogo 658-8501, Japan \label{Konan} \and
RIKEN, 2-1 Hirosawa, Wako, Saitama 351-0198, Japan \label{RIKKEN}
}

\offprints{H.E.S.S.~collaboration,
    \protect\\\email{\href{mailto:contact.hess@hess-experiment.eu}{contact.hess@hess-experiment.eu}};
    \protect\\\protect\footnotemark[1] Corresponding authors
    }

\abstract
{The colliding-wind binary system $\eta$ Carinae has been identified as a source of high-energy (HE, below $\sim$100\,GeV) and very-high-energy (VHE, above $\sim$100\,GeV) gamma rays in the last decade, making it unique among these systems. With its eccentric 5.5-year-long orbit, the periastron passage, during which the stars are separated by only $1-2$\,au, is an intriguing time interval to probe particle acceleration processes within the system. In this work, we report on an extensive VHE observation campaign that for the first time covers the full periastron passage carried out with the High Energy Stereoscopic System (H.E.S.S.) in its 5-telescope configuration with upgraded cameras. VHE gamma-ray emission from $\eta$ Carinae was detected during the periastron passage with a steep spectrum with spectral index $\Gamma= 3.3 \pm 0.2_{\mathrm{stat}} \, \pm 0.1_{\mathrm{syst}}$. Together with previous and follow-up observations, we derive a long-term light curve sampling one full orbit, showing hints of an increase of the VHE flux towards periastron, but no hint of variability during the passage itself. An analysis of contemporaneous Fermi-LAT data shows that the VHE spectrum represents a smooth continuation of the HE spectrum. From modelling the combined spectrum we conclude that the gamma-ray emission region is located at distances of ${\sim}10 - 20$\,au from the centre of mass of the system and that protons are accelerated up to energies of at least several TeV inside the system in this phase.      

}


\authorrunning{H.E.S.S. Collaboration}

\titlerunning{Detection of very-high-energy gamma-ray emission from Eta Carinae during its 2020 periastron passage}
\maketitle

\section{Introduction}

The binary star \etacare (henceforth \etacar) is an extraordinary colliding-wind binary (CWB) system, located in the Carina Nebula, a massive H II region with ongoing star formation at a distance of ${\sim}2.3$~kpc \citep{Smith_2006}. The \etacar\ system is thought to consist of a luminous blue variable with a mass of ${\sim}90 \UNITS{M_{\odot}}$ and a lighter companion with ${\sim}30 \UNITS{M_{\odot}}$ \citep{Hillier2001}.
The system itself is a very prominent source in several wavelength bands partly due to its eruptive history \citep[see e.g.][]{Davidson2012}. It is nowadays widely accepted as one of the few gamma-ray-detected CWB systems with non-thermal emission probed by high-energy X-ray \citep{Hamaguchi2018} and gamma-ray telescopes \citep{fermieta2010,hessetacar}.
In these unique laboratories, particles are accelerated in the wind collision region (WCR) via diffusive shock acceleration \citep[e.g.][]{Eichler1993,Reimer06}. In the WCR, the supersonic winds from the two stars collide to form a pair of standing shocks, separated by a contact discontinuity \citep{Bednarek11}. The wind from the companion (\etacar-B) is fast with a terminal velocity of $v_{\infty} = 3000 \UNITS{km\,s^{-1}}$ \citep{EtaCar:Pittard02} and an associated mass-loss rate of $\dot{M} = 1.4 \times 10^{-5} \UNITS{M_{\odot}\,yr^{-1}}$ \citep{EtaCar:Parkin09}. The primary star's (\etacar-A) wind is comparably slower with $v_{\infty} = 500 \UNITS{km\,s^{-1}}$ \citep{EtaCar:Pittard02}, but denser with $\dot{M} = 5 \times 10^{-4} \UNITS{M_{\odot}\,yr^{-1}}$ \citep{EtaCar:Parkin09}. There is no evidence that the properties of the winds of \etacar-A \citep[see also][]{Groh2012} or \etacar-B have varied over the past 20 years \citep{Gull2022}.

The system is tracked with instruments in several wavelength bands over its full $\sim$5.5-year \citep{Teodoro2016} orbit characterised by a large eccentricity \citep[$\epsilon \approx 0.9$,][]{Mehner_et_al2015}. The periastron passage, at which the stars are separated by only $1-2$ au is linked to strong variability at some wavelengths, such as discovered with spectroscopic measurements of high excitation lines \citep[e.g.][]{Teodoro2016} or with X-ray observations \citep[e.g.][]{Kashi2021}. Variability related to the orbit of the binary was also discovered in non-thermal X-rays \citep{Hamaguchi2018}.

High-energy (HE, $\sim$100\,MeV to $\sim$100\,GeV) gamma-ray emission coincident with \etacar\  was detected shortly after the launch of the \emph{Fermi} satellite by its Large Area Telescope \citep[LAT,][]{fermieta2010} after the initial detection by the AGILE satellite \citep{Tavani2009}. Since then, \etacar\ has been continuously monitored in this energy regime, presently covering 3 periastron passages \citep[see e.g.][]{Reitberger2012, Reitberger2015, Balbo2017, White2020, MartiDevesa21}. The variability in the HE gamma-ray regime, is not as strong as in X-rays, showing only a mild flux increase towards the periastron passage \citep{MartiDevesa21}. The HE spectrum can be described by two distinct components, below and above $\sim$10\,GeV \citep[see e.g.][]{Balbo2017,White2020,MartiDevesa21}. The higher energy component extends beyond the energy range probed by \fermi\ making very-high-energy (VHE, $\sim$100\,GeV to $\sim$100\,TeV) observations crucial to describe the spectrum fully.  Emission at VHE energies from \etacar\ was reported by the High Energy Stereoscopic System (\hess) \citep{hessetacar} based on observations taken before and after the 2014 periastron passage, making it unique among CWBs. 

The periastron passage is of special interest due to the fast-changing conditions. While the thermal X-ray light curve provides crucial information on the shock physics \citep[see e.g.][]{Gull2021,Parkin_et_al_2011}, the combined HE and VHE behaviour is thought to directly probe the acceleration of protons and other nuclei, which have been argued to dominate the emission above 10 GeV \citep[see for example][]{EtaCar:Farnier11,Ohm2015}. Thus, models of cosmic-ray acceleration at shock waves can be tested with VHE gamma-ray measurements.
As such emission can be affected by gamma-gamma absorption in the intense photon fields of the two stars, even on scales of up to $100$\,au \citep{Ohm2015,Steinmassl23}, the shape of the VHE light curve and spectrum can constrain the location of the emission region as well as the underlying emitting particle spectrum.

 \hess\ observed the 2009 periastron passage before the \fermi\ discovery with only limited exposure yielding no detection \citep{hessetacar12}. Unfortunately, the 2014 periastron passage happened outside the visibility period for \hess\ This restricted observations to several months before and after the periastron passage for which a significant VHE gamma-ray signal from \etacar\ was reported \citep{hessetacar}. The 2020 periastron \footnote{Using the period as well as the epoch from \cite{Teodoro2016} the exact passage happened on MJD 58896.6 or February 17, 2020.} was the first periastron passage visible for \hess\ in its 5-telescope array state. Thus, a dedicated and in-depth observation campaign was carried out. 
 
 In this work, we present the results of this campaign together with datasets from previous years and contemporaneous \fermi\ data. In \autoref{sec:hess} the dataset, analysis methods as well as the results of the \hess\ analysis are presented. In \autoref{sec:fermi} the analysis of contemporaneous \fermi\ data is described, before in \autoref{sec:lc} a combined spectrum during the periastron passage is discussed together with both the \fermi\ and \hess\ light curves. These findings are then interpreted in the framework of an existing diffusive shock acceleration model for \etacar\ in \autoref{sec:int}. The main results are then summarised in \autoref{sec:concl} and further conclusions presented.
 
\section{\hess\ data set and data analysis}
\label{sec:hess}
\begin{table*}
\caption[Basic properties of \etacar\ datasets]{Basic properties of \etacar\ datasets}
\centering
\begin{tabular}{l|llllll}

Dataset  & Start Date & End Date & Phase Range  & Live Time &Mean zenith & Telescopes considered\\
& & & & [h] & [ \grad\ ] \\
\hline \\
DS-A & Jan 13, 2013 & May 30, 2016  & $- 0.28$-0.29 & 20.6 & 39.2 & CT1-4\tablefootmark{a} \\
DS-B & Jan 29, 2017 & Apr 7, 2019 & 0.45-0.84 & 63.3 & 39.2 & CT1-4 \tablefootmark{b}\\
DS-C & Dec 23, 2019 & May 24, 2020 & 0.97-1.05 & 97.8 & 39.9 & CT1-5 \tablefootmark{c}\\
DS-D & Feb 15, 2021 & Apr 10, 2021 & 1.18-1.21 & 31.5 & 37.2 & CT1-4 \tablefootmark{b}\\
\hline
\end{tabular}
\tablefoot{The phase is given with respect to the 2014 periastron passage (phase = 0) following the ephemeris from \citet{Teodoro2016}. The live time is given for the stereo observation time and is corrected for dead time.
\tablefoottext{a}{CT1-4 with the original cameras.} 
\tablefoottext{b}{CT1-4 with upgraded cameras \citep[\hess 1U,][]{HESS1U}.}
\tablefoottext{c}{\hess 1U and CT5 with FlashCam \citep{ICRC21FCTechnical}.}}
\label{tab:eta_data}
\end{table*}

 \hess\ is an array of five Imaging Atmospheric Cherenkov Telescopes (IACTs) located in the Khomas
Highland, Namibia. It consists of four 12-m diameter telescopes (CT1-4) upgraded with new cameras in 2016 \citep[\hess1U,][]{HESS1U} and one large central 28-m telescope (CT5) with a new FlashCam camera installed in 2019 \citep{ICRC21FCTechnical,ICRC21FCPerf}. The instrument is sensitive to gamma-rays in an energy range between several tens of GeV and several tens of TeV, depending on observation conditions and analysis settings.

\begin{figure}
\centering
\includegraphics[width = 0.49\textwidth]{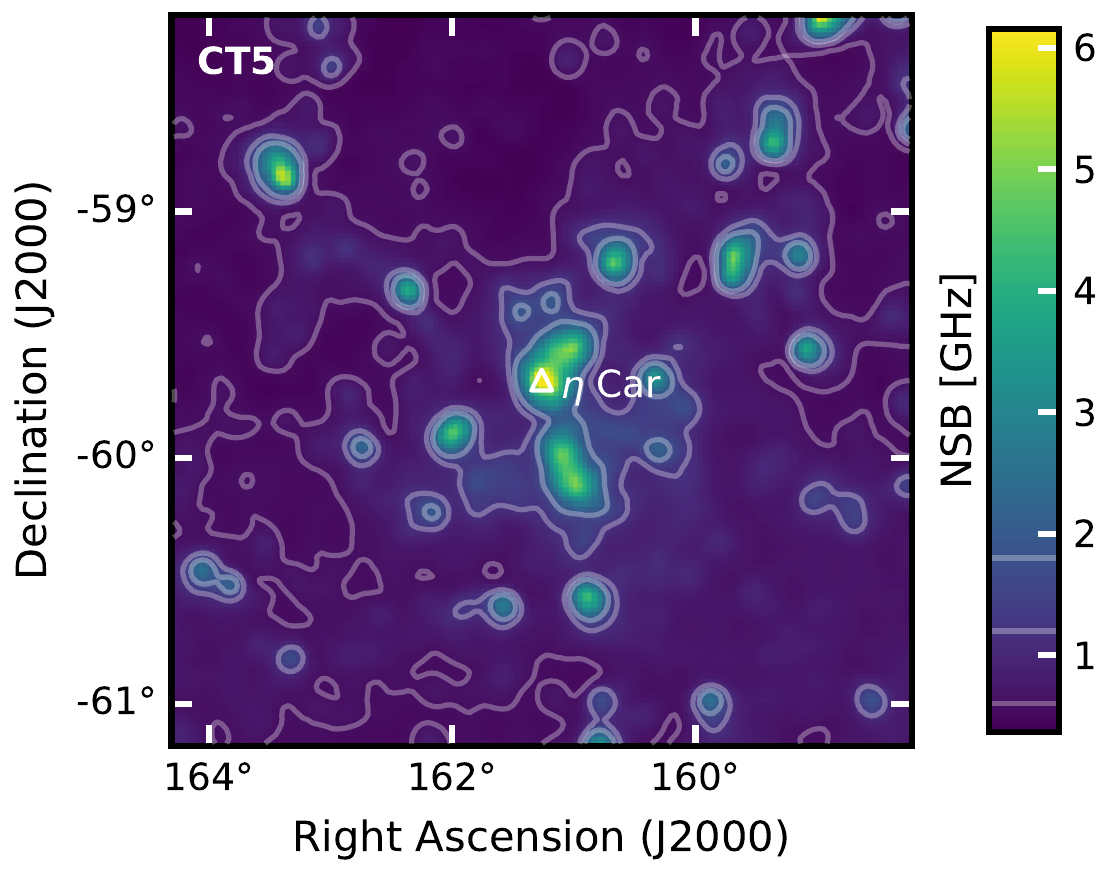}
\caption[]{Average NSB photon rate for the \etacar\ field using CT5.
The map has been derived by averaging the NSB maps from all observations of the 2020 dataset (DS-C). The position of \etacar\ is shown with a white triangle. Contour levels for better visualisation have been added at 0.6, 1.2 and 1.8 GHz. The map is zoomed in from the full FoV for better visibility of the main features.}
\label{fig:eta_nsbmap}
\end{figure}

The strategy for gamma-ray observations of \etacar\ with the updated \hess\, array was customised due to the unique field of view (FoV). \etacar\ is situated at the heart of the Carina Nebula, which appears in optical wavelengths as a region of large-scale diffuse emission. 
This increases the night sky background (NSB) photon rate and hence the noise level for IACT observations, which are based on signals recorded by photomultipliers sensitive in the optical and near-ultraviolet waveband. The NSB can be estimated either using the baseline shift of each event \citep[for FlashCam, ][]{Flashcam17} or the width of the pedestal distribution \citep[for CT1-4,][]{HESSCalib2004}. For each telescope, the average NSB during \etacar\ observations is several times the value typically reached in observations along the Galactic plane, which is $\sim$300 MHz for CT5. As presented in \autoref{fig:eta_nsbmap} the NSB rate reaches more than 10 times this value in small regions of the FoV. To allow for data taking with stable trigger rates and small dead time an increased trigger threshold for all five H.E.S.S. telescopes was used. For CT1-4 the pixel trigger threshold was increased from 5.5 photo-electrons (p.e.) to 6.5 p.e for the \hess1U observations, which needs to be fulfilled for at least 3 pixels. For FlashCam the bright-source trigger threshold of 91 p.e. (instead of 69 p.e.) per 9-pixel sum as defined in \cite{ICRC21FCTechnical} was used. These trigger settings were derived using a scan on the \etacar\ field itself and are chosen to prevent unstable trigger behaviour induced by random noise triggers. This is particularly important for CT5 which is operated with a monoscopic trigger and is further analysed utilising a monoscopic analysis \citep{Murach2015} in this work.

Furthermore, the gain of the photomultipliers in the FlashCam camera on CT5 was reduced to the value otherwise used for observations during bright moonlight \citep{ICRC21FCTechnical}. Consequently, the number of switched-off pixels to prevent high anode currents, even though it is still considerably larger than for other observation targets, could be reduced to less than 3 \% of all pixels in each camera.

\etacar\ is visible for \hess\ at moderate zenith angles starting from December until the beginning of June each year. The adopted observation strategy was to obtain observations at zenith angles lower than $40$\grad\ or $45$\grad\ depending on the dataset. All data were taken using the wobble mode observation technique \citep{CrabObs} with a typical pointing offset of 0.7$^\circ$ from the source position.  During the period around the actual periastron passage in February 2020, a loosened zenith angle requirement of up to 60\grad\ was used to maximise the exposure during this period. Unfortunately, the available observation time, especially in February, was limited due to the Namibian rainy season. Further data were acquired in 2021, assuring sufficient coverage over the entire 5.5-year orbital period together with the existing data from previous years. 

A run\footnote{A \hess\ run is an observation entity that lasts by default 1680 seconds.} selection based on the standard spectral selection criteria \citep{CrabObs} could be applied for the CT1-4 data in the 2013-2016 time period (phase\footnote{The ephemeris from \citep{Teodoro2016} was also used for the definitions of phase in this work. The phases of the datasets at hand are quoted with respect to the 2014 periastron passage (phase = 0)}: $-0.28$-0.29, DS-A), as there the trigger threshold was still set at its default value. This yielded 20.6 h of good-quality data in this time period, whereas the combined observation campaigns of 2017 to 2019 (phase: 0.45-0.84, DS-B) had a live time of 63.3 h after selection. For the data taken after 2019 a run selection based on trigger rate stability (RMS $\leq 50$~Hz for CT5) and run duration ($\geq 1500$~s)  was used. This resulted in a final dataset of 97.8 h during the 2020 periastron campaign (phase 0.97 to 1.05, DS-C) and 31.5 h in 2021 (phase: 1.18-1.21, DS-D). For DS-D only stereo data is considered due to the limited dataset and the large systematics of the mono analysis. The basic properties of the datasets are summarised in \autoref{tab:eta_data}.

\subsection{CT5 mono analysis}
\label{sec:hess_mono}

The data from the large central telescope can be analysed with the monoscopic analysis method.
As this method only considers data from one telescope both the energy and the direction reconstruction are computed utilising multi-layered perceptrons with the image parameters of cleaned simulation images as input for the training following \citet{Murach2015}.
The exceptional noise levels in the \etacar\ observations result in a mismatch with simulations that assume standard observation conditions. Therefore, a designated simulation set taking into account the actually measured NSB (see \autoref{fig:eta_nsbmap}) has been produced with the CORSIKA \citep{CorsikaKIT} and sim\_telarray \citep{simtel2008} packages. This simulation set was then utilised to produce custom instrument response functions valid for the \etacar\ observations. This involved retraining of the neural networks\footnote{The neural networks were trained with the machine learning open source platform {\it TensorFlow} using the {\it Keras} python interface} used for reconstruction as well as for the gamma-hadron separation. The cuts for the gamma-hadron optimisation and the integration radius were optimised on simulations and real events from source-free FoVs with increased noise level for a source with a steep spectral index (3.7) and a flux of 2 \% of the Crab Nebula \citep{CrabObs}. 

To get an accurate description of the background and its acceptance, test runs targeted at other sources were transformed to mimic the characteristics of reference runs from the \etacar\ field. The transformation was applied such that the test runs would emulate the reference \etacar\ run in pointing, time, and camera features such as the measured pedestal width and broken pixel pattern \citep{HESSCalib2004}. To properly imitate the influence of increased and inhomogeneous noise, additional noise was added on a per-pixel basis to match the noise level of the reference run. This procedure was repeated for every \etacar\ run.  These runs were then stacked and utilised through an On-Off background approach \citep{Berge2007} using the Gammapy (version 1.0) software package \citep{gammapy2023,gammapy:zenodo}. 
An additional cut on the shower displacement variable which describes the offset from the reconstructed shower position to the shower's centre of gravity \citep[see][for definition]{Murach2015} was introduced. Showers influenced by NSB and broken pixels, as is the case in the central region of the Carina Nebula, are more likely to be incorrectly reconstructed close to their own centre of gravity. Therefore, a lower limit on the shower displacement variable was introduced, as described in more detail in \autoref{app:delta}, to reduce the number of showers severely affected by these effects.

\begin{figure}
\centering
\includegraphics[width = 0.24\textwidth]{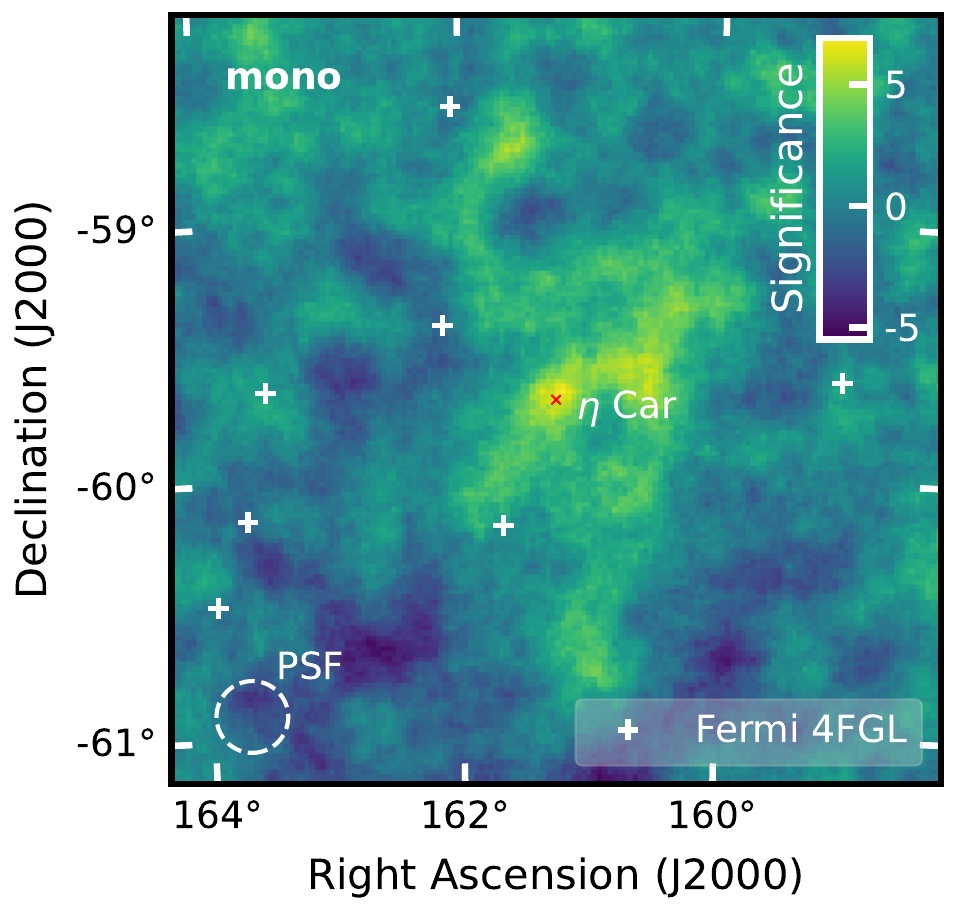}
\includegraphics[width = 0.24\textwidth]{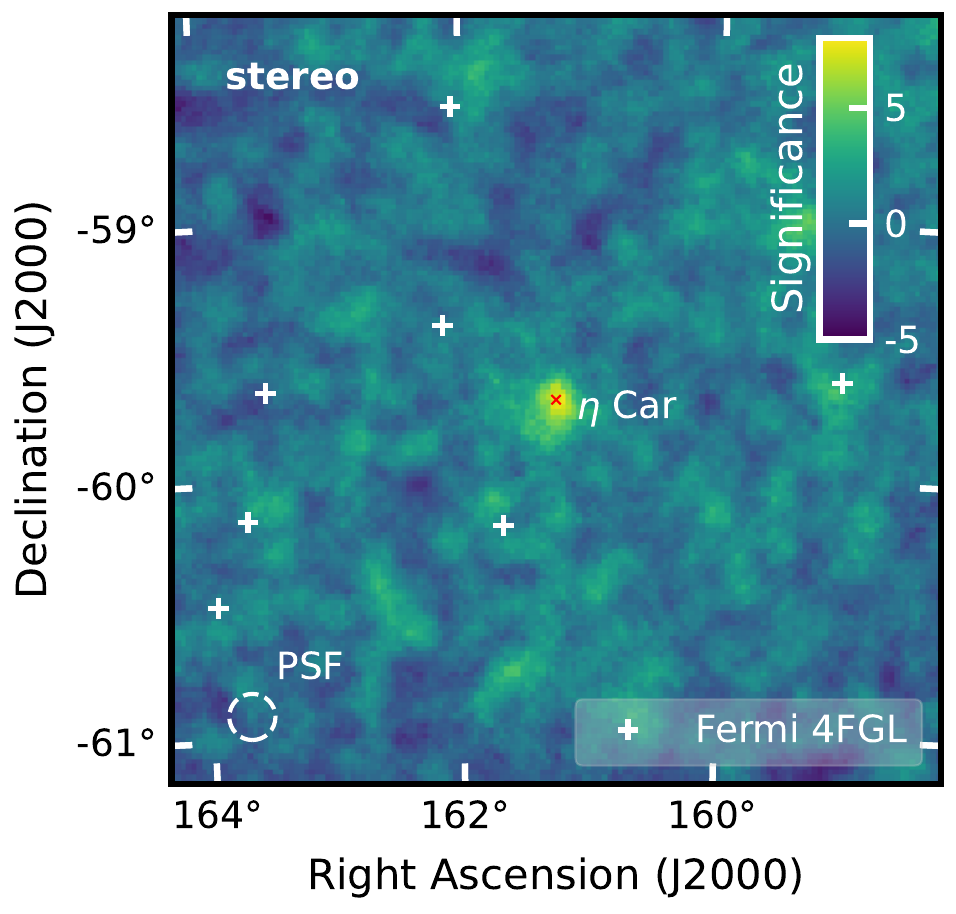}
\includegraphics[width = 0.24\textwidth]{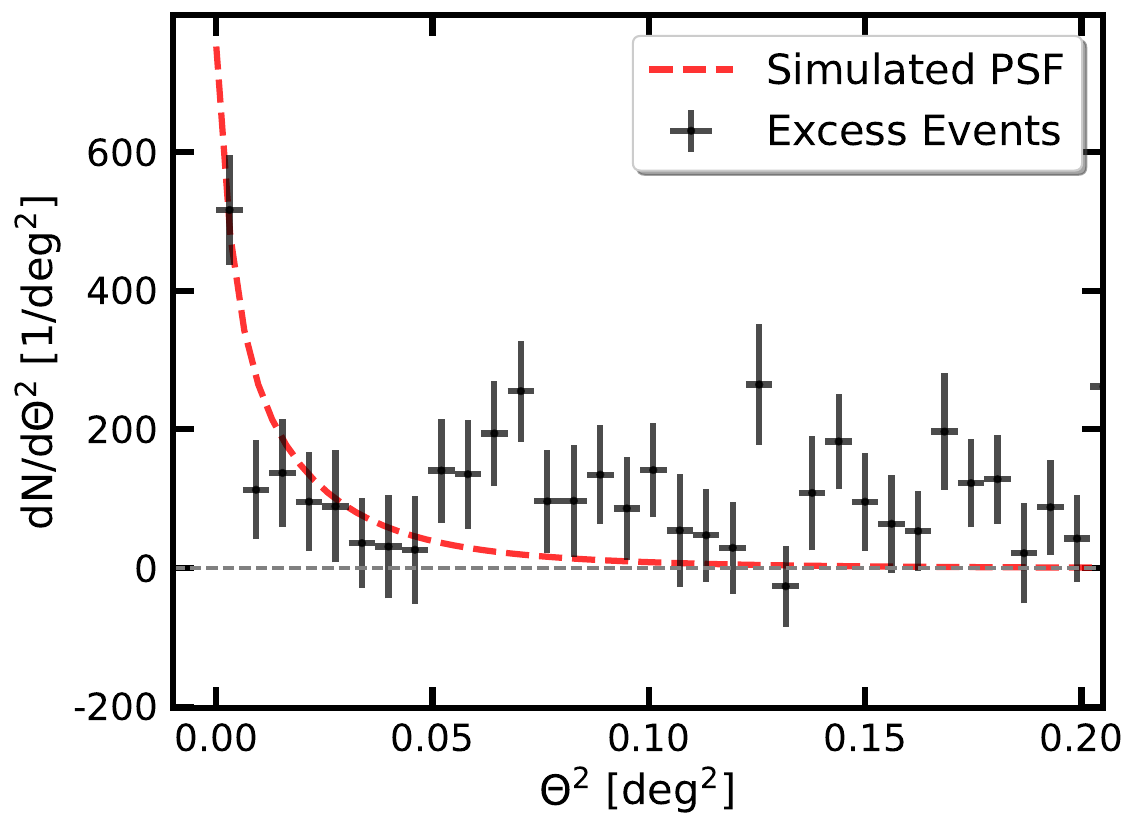}
\includegraphics[width = 0.24\textwidth]
{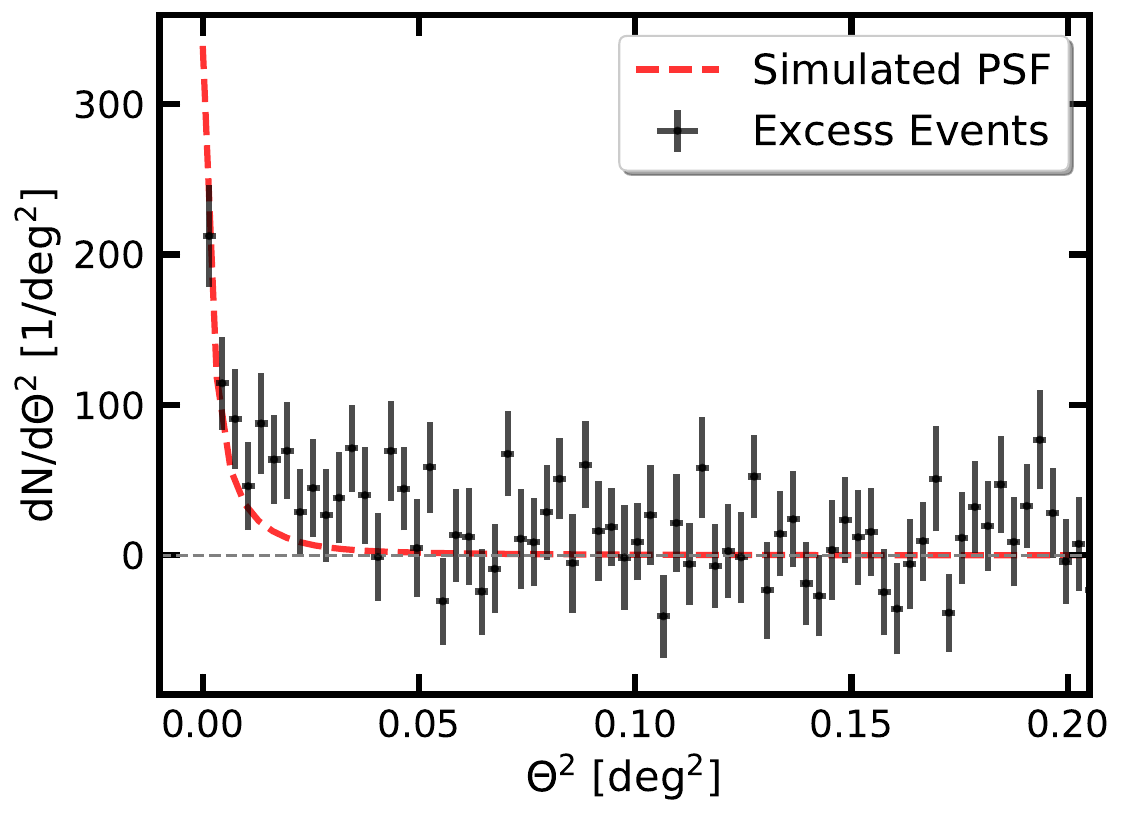}
\caption{Significance maps and radial excess distributions of the \etacar\ periastron dataset (DS-C).
The resulting significance maps for CT5 mono (upper left) and CT1-4 stereo (upper right) are shown with correlation radii of 0.13\grad\ and 0.09\grad , respectively. The position of \etacar\ is marked with a red cross. Additionally, all other \fermi\ 4FGL sources \citep{4fgldr3} in the FoV are marked. The radial excess distributions for mono (lower left) and stereo (lower right) analyses are centred on the \etacar\ position. They are compared to the PSF derived from simulations with an assumed index of 3.3. The scale
of the PSF was matched to the first bin.}
\label{fig:eta_resultmaps_sig}
\end{figure}

The resulting significance map for the 2020 dataset (DS-C) with a correlation radius of 0.13\grad\ is shown in \autoref{fig:eta_resultmaps_sig}. At the \etacar\ position, a significance of 7.1 $\sigma$ is found using the approach described in \citet{LiMa1983}. The signal appears point-like and the residual significance distribution is described by a Gaussian with a width of 1.3, wider than the desired case (see \autoref{sec:app-h}). 

A circular region with size $\Theta = 0.13$\grad\ centred at \etacar , the result of the cut-optimisation procedure, was chosen to derive the spectral properties modelled with a power law of the form $\phi = \phi_0 \left( \frac{E}{E_0} \right) ^{-\Gamma}$. The lower energy threshold of the analysis, defined as the energy bin at which 10\,\% of the maximal effective area is reached, is 0.14 TeV. The fit resulted in a soft spectrum with a spectral index of  $\Gamma= 3.3 \pm 0.4_{\mathrm{stat}}\, \pm 0.2_{\mathrm{syst}}$ and a flux normalisation of $\phi_0 = (4.5 \pm 1.0_{\mathrm{stat}}\, ^{+0.9} _{-2.3\,\mathrm{syst}}) \times 10^{-11} \UNITS{TeV^{-1}\,cm^{-2}\,s^{-1}}$ at a fixed reference energy $E_0$ of 0.2 TeV. The spectrum is presented in \autoref{fig:eta_spectralhess}.

We note that the spectral properties derived with the customised analysis configuration for DS-C are inconsistent with the results presented in \citet{hessetacar} based on observations around the 2014 periastron passage.
The differences are attributed to improvements in the understanding of noise levels and the effects on data analysis when using FlashCam. Systematics were underestimated in the results presented previously due to the impact of the extreme NSB in the \etacar\ FoV on the mono analysis method (for more details see \autoref{sec:app-h}).
Consequently, no physics conclusions, especially concerning variability, should be drawn from comparison with the previous work.

We adopted a conservative treatment of the systematics of the spectral parameters, in particular for the flux normalisation. For previous mono analyses with FlashCam data, a systematic uncertainty on the flux normalisation of ${\sim}20$\,\% \citep[e.g.][]{Novahess2022} was estimated. Due to the reduced, but still remaining influence of NSB the systematic uncertainty for the analyses presented in this work can be assumed to be asymmetric, potentially biasing the result towards a higher flux. This is evident by the wider-than-expected residual significance distribution and the remaining features in the significance map, which can not be associated with sources from the \fermi\ 4FGL \citep{4fgldr3} catalogue. 
Therefore, it can be assumed that these are remaining noise artefacts.  
From the significance of the remaining features in the significance map, a potential residual reconstructed emission caused by NSB translating into a systematic uncertainty of the flux level of ${\sim} 50$\,\% can be estimated.

\subsection{CT1-4 stereo analysis}

For the CT1-4 stereo analysis, the ImPACT reconstruction technique \citep{Parsons2014} was chosen. As the stereo direction reconstruction is more robust against the effects of noise and the noise level is taken into account for the ImPACT-based reconstruction, no custom analysis configuration and background modelling method had to be employed. Instead, the ring background method and the reflected regions background method \citep{Berge2007} were utilised for the derivation of maps and spectra, respectively. 
The analysis was reproduced with a second independent reconstruction method \citep{HapFr}. This cross-check analysis yielded compatible results within uncertainties.

For the 2020 periastron dataset (DS-C) a point-like signal with a significance of 8.5 $\sigma$ is found at the \etacar\ position using the default integration radius of 0.07\grad . The resulting significance map is shown in \autoref{fig:eta_resultmaps_sig}. If compared to other \fermi\ 4FGL sources \citep{4fgldr3}, no other sources are significantly detected.
The background outside of the exclusion region is normalised with a significance distribution fitted by a Gaussian with a width of 1.17 and a mean of -0.02, also a bit wider than desired.

\begin{figure}
\centering
\includegraphics[width = 0.49\textwidth]{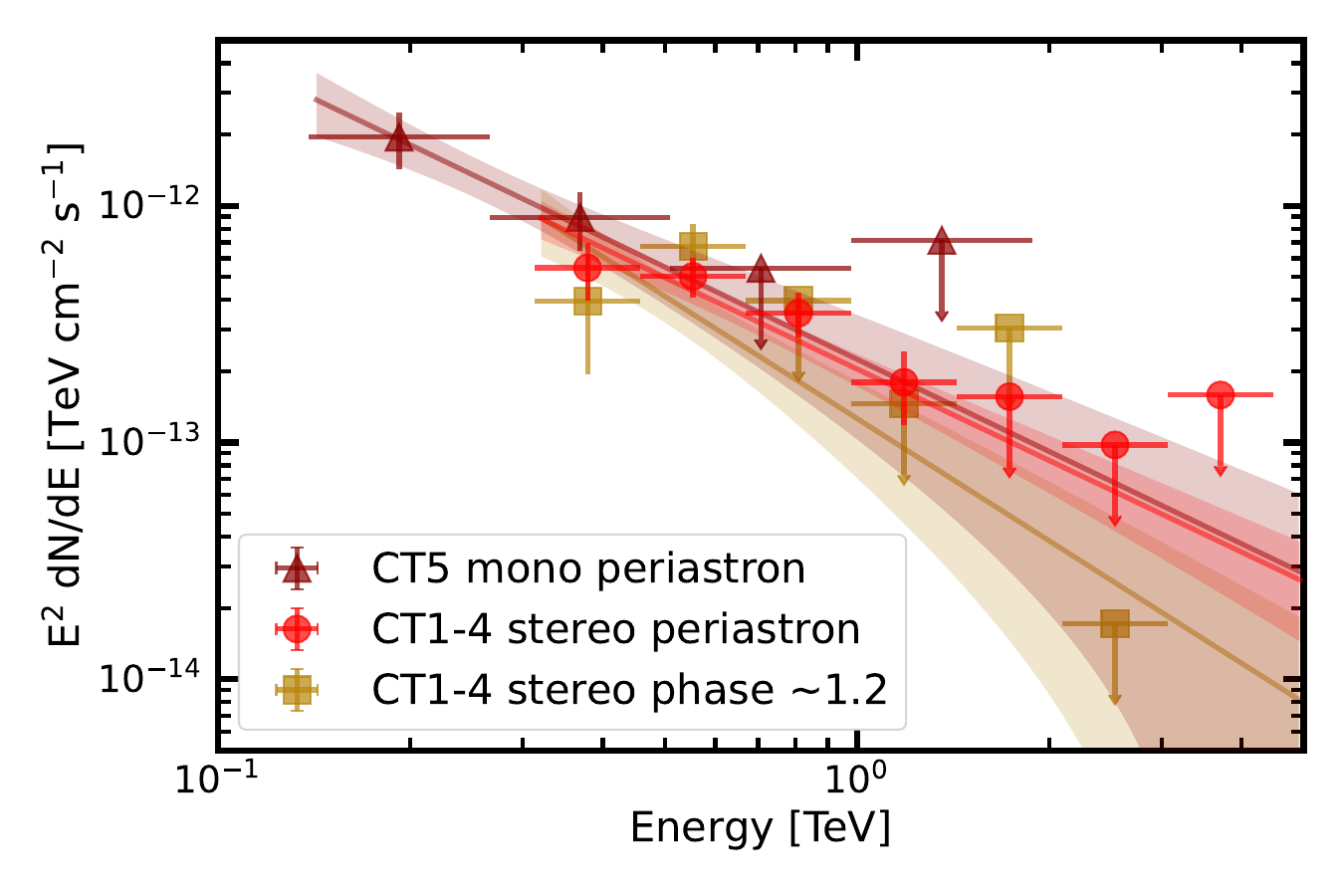}

\caption{
Resulting spectral model from the stereo analysis of the periastron dataset (DS-C, red circles) compared to the spectral model derived in the mono analysis (dark red triangles) of the same dataset. Furthermore, the best-fit spectral model for the 2021 dataset (DS-D, yellow squares) at phase ${\sim} 1.2$ is shown. Additionally, also the derived flux points are compared. The error estimates plotted are statistical uncertainties only.  
}
\label{fig:eta_spectralhess}
\end{figure}

The lower energy threshold is 0.31 TeV and the reference energy was fixed at 1 TeV.
Assuming again a power-law spectrum, the best-fit result yielded a spectral index of  $\Gamma= 3.3 \pm 0.2_{\mathrm{stat}} \, \pm 0.1_{\mathrm{syst}}$, and a flux normalisation of $\phi_0 = (2.0 \pm 0.3_{\mathrm{stat}} \, \pm 0.5_{\mathrm{syst}}) \times 10^{-13} \UNITS{TeV^{-1}\,cm^{-2}\,s^{-1} }$, consistent with the mono result, as shown in \autoref{fig:eta_spectralhess}.
The highest significant flux point has an energy of 1.18 TeV and bin edges of 0.98 TeV and 1.43 TeV. In the following, this will be adopted as a tentative estimate of the maximum photon energy. 

The 2021 dataset (DS-D) was analysed with the same methods as DS-C yielding a 5.1 $\sigma$ detection and a consistent spectrum (see as well \autoref{fig:eta_spectralhess}) with a spectral index of $\Gamma= 3.7 \pm \, 0.5_{\mathrm{stat}} \pm 0.1_{\mathrm{syst}}$ and a flux normalisation of $\phi_0 = (1.3 \pm 0.6_{\mathrm{stat}} \pm 0.3_{\mathrm{syst}}) \times 10^{-13} \UNITS{TeV^{-1}\,cm^{-2}\,s^{-1} }$ at a reference energy of 1 TeV.

As the stereo CT1-4 analysis uses standard configurations and methods, the systematic uncertainties were taken to be on a comparable level to \citet{CrabObs} with an additional systematics of 15\% on the flux normalisation based on the larger than expected width of the residual significance distribution. This results in a total systematic error on the index of 0.1 and the flux of 25 \%.

\section{\fermi\ data analysis}
\label{sec:fermi}
 Data from \fermi\ was analysed in a time interval contemporaneous to the \hess\ periastron dataset (DS-C). In order to retrieve a steady model of the FoV, in a first iteration a fit was performed on the full 14-year dataset of \fermi\ on that region. The data selection was based on the latest \fermi\ Pass 8 data \citep{Bruel2018} starting from Aug 4$^{\rm th}$, 2008 (MET 239557417) to Oct 26$^{\rm th}$, 2022 (MET 688521600). Events over an energy range from 100 MeV to 500 GeV were included from a region of interest (ROI) of \SI{20}{\degree} by \SI{20}{\degree}, centred at the nominal position of \etacar\ and using equally spaced binning (0.1\grad) in Galactic coordinates. Data were selected using the SOURCE event class ({\it evclass=128}) with joint FRONT+BACK event types ({\it evtype=3}). Time intervals in which the ROI was observed at a zenith angle greater than \SI{90}{\degree} were discarded. Data were analysed using the Fermitools (version 2.2.0), which is the official {\it ScienceTools} suite provided by the Fermi Science Support Center\footnote{\url{https://fermi.gsfc.nasa.gov/ssc/}} and FermiPy (version 1.2) \citep{Fermipy2017} software packages. 
 
 The model of sources surrounding \etacar\ includes all sources in the ROI from the \emph{Fermi}-LAT 12-year source catalogue \citep[4FGL-DR3,][]{4fgldr3}, except the unidentified point source 4FGL J1046.7-6010. This particular source is located in the centre of the molecular cloud Southern Pillars \citep{Rebolledo2015}, which was added as a diffuse source together with other molecular cloud templates as described in  \citet{Steinmassl23}. The exact analysis details are summarised in \autoref{tab:fermi_cfg}.
 
 After retrieving a model valid for the full time interval, all data between the start (MET 598755697) and the end (MET 612045512) of the \hess\ periastron dataset (see DS-C in \autoref{tab:eta_data}) were selected. The spectral properties of \etacar\ (4FGL J1045.1-5940), modelled as a point source with a log-parabola spectrum $\phi=\phi_0(E/E_0)^{-\alpha-\beta\log(E/E_0)}$, were freed and fit. Next, sources associated with large residuals (test statistic\footnote{The test statistic is defined as the difference in the Cash statistic \citep{Cash79} between the models with and without the evaluated source.} $\Delta$TS > 25)  in the periastron dataset were additionally freed and the fit was repeated to improve the overall fit quality. Using this approach a highly significant excess from the position of \etacar\ ($\Delta$TS = 489) was found. The retrieved best-fit model for \etacar , with $\alpha = 2.31 \pm 0.08$, $\beta = 0.18 \pm 0.04$, $\phi_0 = (2.99 \pm 0.23) \times 10^{-6}  \, \UNITS{cm^{-2}\,erg^{-1}\,s^{-1}}$  at a fixed reference energy $E_0 = 2.11 \, \UNITS{GeV}$, is consistent with previous studies \citep{MartiDevesa21,Steinmassl23}.
 
 Furthermore, from the analysis of the full-time interval a temporal analysis was obtained using the {\it lightcurve} function of FermiPy. For this, time bins of $\sim 60$ days (phase = 0.03) were defined and the model parameters of \etacar\ and the isotropic background component were freed.

\section{Combined spectrum and light curve}
\label{sec:lc}
\begin{figure}
\centering
\includegraphics[width = 0.49\textwidth]{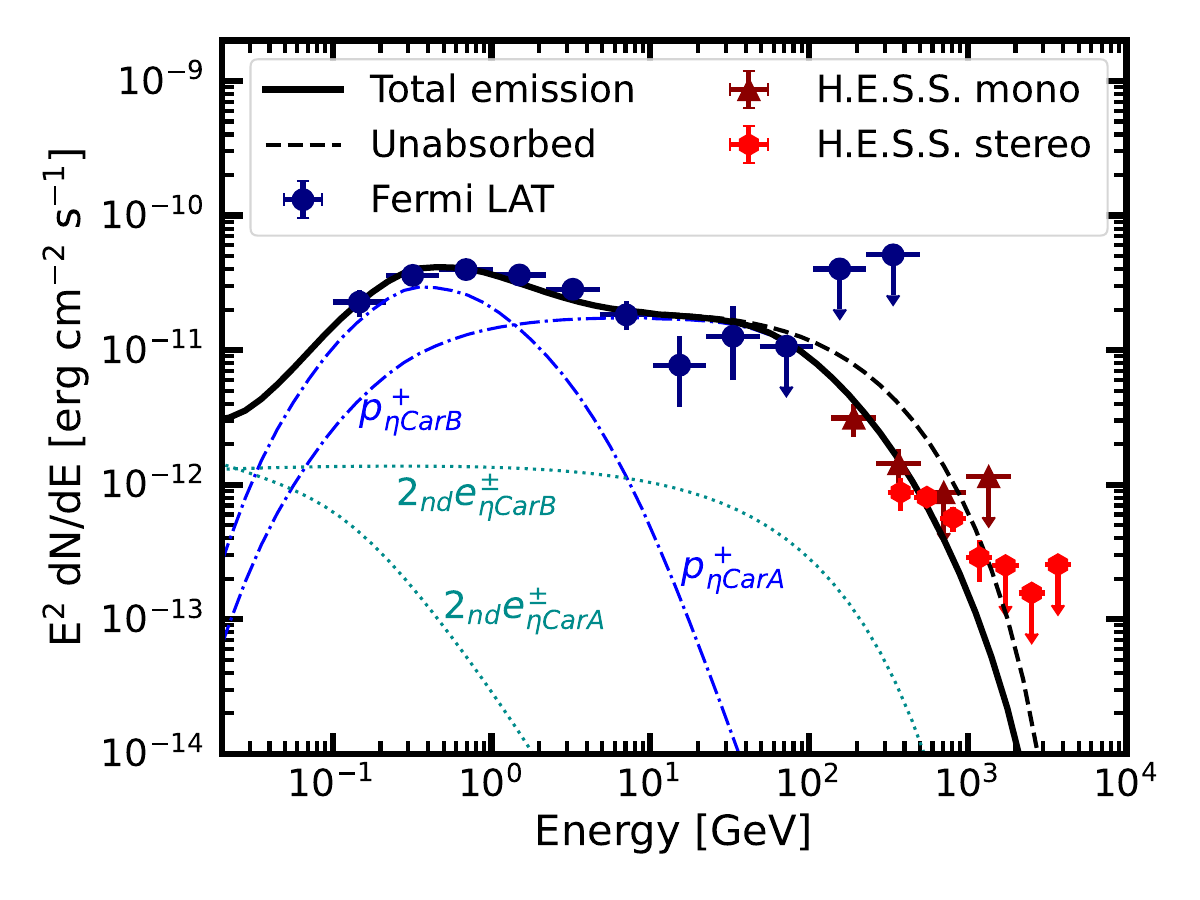}
\caption{\fermi\ and \hess\ flux points for the SED obtained around the 2020 periastron (orbital phase 0.97-1.05). The total periastron model as well as the individual components from \citet{White2020}, assuming a radial distance of the gamma-ray emission of 20~au as discussed in the text, is shown. The individual components denote the absorbed hadronic and secondary leptonic components from both winds. The dashed line denotes the unabsorbed model.}
\label{fig:eta_model_20au}
\end{figure}

The spectral result for the \hess\ periastron observation corresponds to the phase interval from 0.97 to 1.05. Overall, the combined spectral energy distribution (SED)  of the \hess\ and the \fermi\ flux points extends from 100 MeV up to TeV energies (see \autoref{fig:eta_model_20au}). 
Its shape highlights the two components (below and above ${\sim}$ 10 GeV) already revealed by other studies \citep[e.g.][]{EtaCar:Farnier11,Balbo2017,MartiDevesa21}.

\begin{figure}
\centering
\includegraphics[width = 0.49\textwidth]{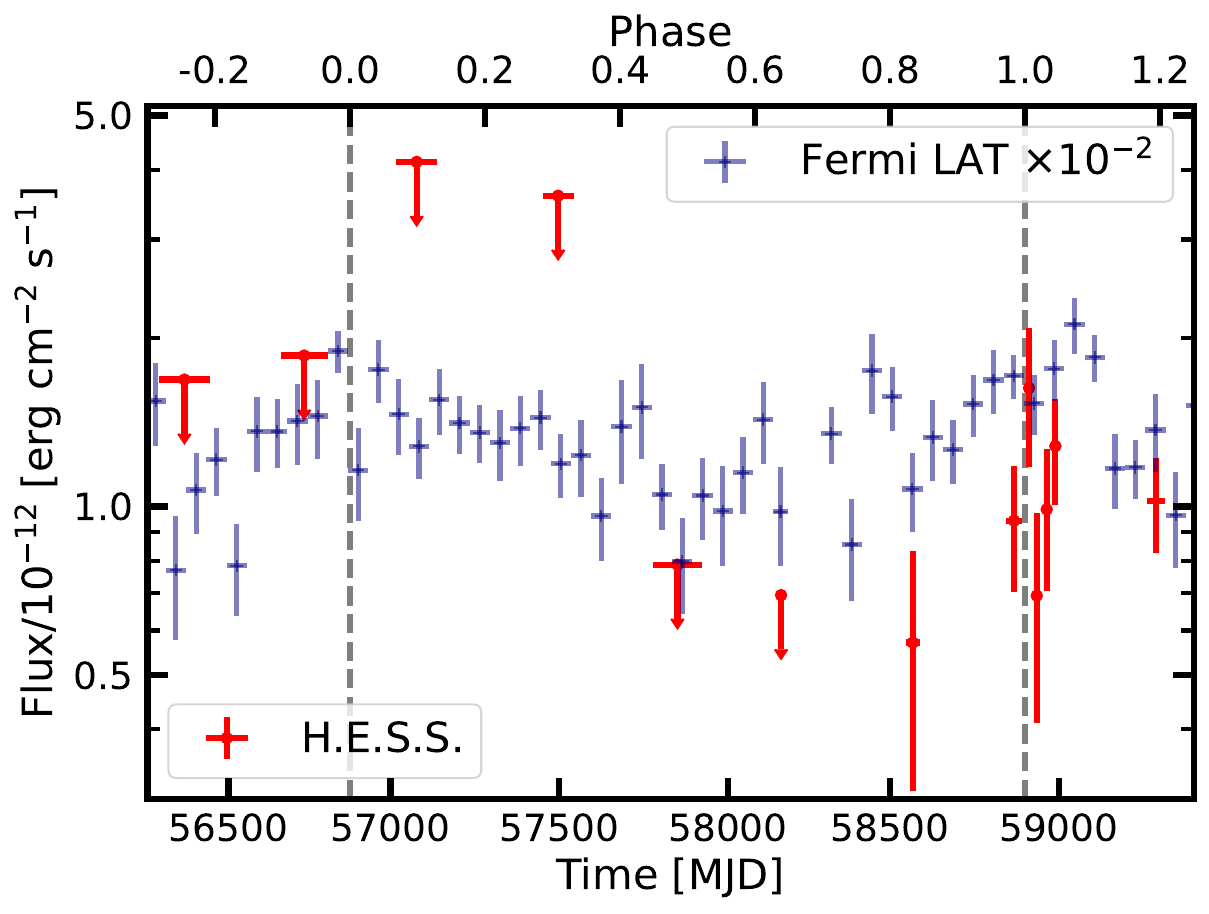}
\caption{
The \hess\ light curve (red) above 310 GeV derived from the stereo data in this work is compared to the \fermi\ light curve (blue) above 100 MeV (scaled down with $10^{-2}$) with a bin size of $\sim 60$ days. 
The dashed lines represent the periastron passages of 2014 and 2020 at phases 0 and 1, respectively. The error bars correspond to the 68 \% confidence interval and upper limits (at 95 \% confidence) are calculated for phase bins with less than 2 $\sigma$ detection significance.}
\label{fig:eta_lc_multi}
\end{figure}

The datasets for the years from 2013 to 2019 (see \autoref{tab:eta_data}) were also analysed with the reflected regions background method to derive a long-term light curve (see \autoref{fig:eta_lc_multi}). As the systematics of the mono reconstruction on the \etacar\ field are large (\autoref{sec:hess_mono}), the light curve is based on the stereoscopic reconstruction of CT1-4 data only. Stereoscopic datasets from the years before 2012 have already been published in \citet{hessetacar12}. The data-taking stability during the early years of \etacar\ observations was not yet optimised with tailored trigger settings for CT1-4 and during some observation phases, only CT5 mono observations were taken. Therefore, the available CT1-4 dataset is rather limited, especially for the years 2013 to 2016. The stereo data taken around the last periastron was not considered in \citet{hessetacar} but is taken into account in this work. However, only a few runs pass the selection cuts. The data were analysed separately for the DS-A and DS-B datasets (as defined in \autoref{tab:eta_data}). 

In the combined DS-A dataset the total significance at the \etacar\ position was below $2 \sigma$. Hence only upper limits for a light curve could be derived. For this, the time bins were chosen according to the observation campaigns in the different years. The energy flux for the light curve in general was derived between 0.31 and 3.1 TeV.

In the 2017-19 data (DS-B) the detection significance reaches only $3.4 \sigma$. 
For this period a spectral index of $\Gamma= 3.6 \pm 0.5_{\mathrm{stat}}\, \pm 0.1_{\mathrm{syst}}$ and a flux normalisation of $\phi_0 = (0.7 \pm 0.3_{\mathrm{stat}}\, \pm 0.2_{\mathrm{syst}}) \times 10^{-13} \UNITS{TeV^{-1}\,cm^{-2}\,s^{-1} }$ at a reference energy of 1 TeV was retrieved, even though these should be considered only as tentative estimates due to the low significance.  
To obtain light curve points the time bins were again chosen according to the observation campaigns in the different years. With this, only the 2019 data yielded a flux point above the $2 \sigma$ threshold, with a significance of $2.3 \sigma$. 

Due to the large quantity of observational data, the 2020 periastron dataset (DS-C) could be split up into shorter time bins, each of several weeks duration. The bins were defined from the different periods between observation breaks in \hess\ due to the bright moon. The resulting light curve as a function of time and phase can be seen in \autoref{fig:eta_lc_multi}. A reduction of the flux before the 2020 periastron passage in the combined flux of DS-B (phase 0.45 - 0.84) is hinted, whereas the flux rises towards the 2020 periastron. Nevertheless, the null hypothesis of a constant flux cannot be ruled out. During the periastron passage, no clear variability within the uncertainties on short time scales of roughly a month is present. At phase 1.18 - 1.21 (corresponding to DS-D) the flux is still comparable to the periastron flux. If the hint can be confirmed, the general behaviour of the long-term VHE light curve is rather similar to the \fermi\ light curve above 100 MeV derived using the full dataset described in \autoref{sec:fermi}.

\section{Interpretation and discussion}
\label{sec:int}
The emission of the high energy component (above 10 GeV) is generally believed to be of hadronic origin \citep[e.g.][]{EtaCar:Farnier11,Ohm2015,Balbo2017,MartiDevesa21,Steinmassl23}. Producing photons with these energies in a leptonic scenario is challenging. Estimations of the stellar wind/shock parameters imply a Hillas limit of several tens of TeV \citep[see e.g.][]{hessetacar}, well in line with the observed gamma-ray spectrum reaching up to TeV energies.    

At each of the two shocks associated with the two winds \citep{Eichler1993}, protons and electrons are accelerated. To derive a resulting gamma-ray spectrum we employ a colliding wind shock-acceleration model \cite[see][]{White2020}. Within this model, protons accelerated at the shock of the companion star radiating via proton-proton interactions are the main contributing population to VHE gamma-rays. This is similar to other models \citep{Gupta2017}.

The picture in the lower energy gamma-ray domain is not as clear, and the hadronic scenario of the model, assigning the emission below 10 GeV to protons accelerated in the primary wind, is not yet fully confirmed \citep{MartiDevesa21}. Nevertheless, there is a clear hint of a spectral break at energies consistent with the interpretation of a pion bump \citep{2022ApJ...933..204A}. As the focus of this work is on the higher energy domain, the lower energy component is not further discussed.

To match the observed \hess\ data from the 2020 periastron passage, the predicted gamma-ray spectra at the periastron phase before absorption (phase 0.995 to 1.025) were taken from \citet{White2020}. In their model, the highest energy protons are accelerated at the termination shock of the companion's wind, and mix with the shocked material of the denser primary wind. The gamma-ray emission is thus produced in the ballistic region right outside the shock cap \citep[see][]{Parkin2008}, which during periastron is located close to the two luminous stars. The resulting absorption, and associated reduction of the \hess\ flux around periastron, is too strong to explain the findings in this work (see \autoref{sec:app-model} and \autoref{fig:eta_modelabs}).

Hence, a reduced gamma-gamma absorption during periastron with respect to their model needs to be assumed. To reduce the absorption, the emission region of gamma rays should be located further out. This effectively implies an enlarged mixing region in the system. To account for this, transparency curves were calculated at several radial distances from the centre of mass of the binary system \citep[more details can be found in][]{Steinmassl23}. The emission was assumed to be spherically symmetric and the geometry of the system adapted to periastron assuming the stellar and orbital parameters as summarised in \autoref{tab:orbit} and \autoref{tab:whitemodel}.

When compared to the \hess\ data during periastron, the modelled flux level is better matched if the emission takes place at a distance of 10~au to 20~au from the binary. Further details are provided in \autoref{sec:app-model}. In \autoref{fig:eta_model_20au}, the individual components and the total emission of the model are compared to the data assuming a radial distance of 20~au for the gamma-ray emission. For this, the normalisation of the flux from \etacar-A was increased by a small factor (1.1) with respect to \citet{White2020} to better match the observed \fermi\ flux during the 2020 periastron. The \hess\ measurements can be explained by hadronic emission from protons accelerated at the shock of \etacar-B.

Even though the adapted model matches the overall \hess\ flux points well, the \hess\ spectrum hints at a harder spectral behaviour than suggested by the model. To account for that as well, the cut-off energy of the proton spectrum can be varied. We utilised the proton spectrum from \citet{White2020} with a spectral index of 1.8 and a cut-off at 1.1~TeV ( more details in \autoref{sec:app-model}). The proton cut-off energy was varied using the Gamera package \citep{gamera_2015,gamera_2022} to better match the \hess\ flux points, keeping the index and normalisation fixed. 
The resulting gamma-ray SEDs assuming emission from a radial distance of 20~au were compared to the high energy \fermi\ and \hess\ flux points. As expected, a higher cut-off energy ${\sim} 1.5$~TeV describes the shape of the \hess\ SED better  (see as well \autoref{sec:app-model}). As a plausibility check, such a small increase in the cut-off energy for the protons accelerated at the shock of \etacar-B is still in line with the Hillas limit of several tens of TeV \citep{hessetacar} and the otherwise poorly constrained model parameters \citep[see][for details]{White2020}. As the model itself has many free parameters and the information presented by the combined SED is limited, a proper fit is not attempted. 

With these estimates, it can be concluded that within an alternative setup of parameters and location of the emission zone from \citet{White2020} the spectrum of the 2020 periastron passage is well described, if the emission region is moved out to a distance of ${\sim} 20$~au, e.g. through a longer mixing length, and a slightly higher cut-off for the hadronic component accelerated in \etacar-B is assumed. These findings are also in line with the light curve of \etacar\ that shows no flux reduction for the periastron passage, which would be expected for a strong gamma-gamma absorption during periastron. The increase of flux towards periastron yielding also higher maximum photon energies in the periastron SED can be explained by the higher density in the ballistic region during that period making p-p interactions more frequent \citep{Ohm2015}.   

\section{Summary and conclusion}
\label{sec:concl}
In this work, the study of the VHE gamma-ray emission of \etacar\ was presented with special emphasis on the 2020 periastron passage. \etacar\ is situated in a peculiar FoV characterised by large and inhomogeneous NSB, testing the observational limits for IACTs. 

Therefore, special observation and analysis settings were needed to achieve stable data taking. 
For the mono analysis, a very careful treatment of noise factors had to be set up and applied to retrieve robust results. This customised mono analysis approach was specifically designed for regions with and inhomogeneous NSB and could also be valuable for potential other observations of sources with steep spectra (e.g. transients) in bright NSB regions or with bright, inhomogeneous moonlight over the field of view.  

For \etacar\ significant gamma-ray emission was detected independently utilising mono and stereo analysis techniques. Both analyses agree in the spectral shape and flux normalisation, resulting in a periastron SED above energies of 130 GeV described as a steep power law with an index $\Gamma \approx 3.3$. 
The spectral properties derived with both analysis methods are inconsistent in the flux level with the results derived in \citet{hessetacar}. This is attributed to underestimated systematics in the previous work (see \autoref{sec:app-h}). To retrieve the flux at the phases probed by the previous work (phase -0.22 to -0.04 and 0.09 to 0.1 in our phase definition), H.E.S.S. observations around the 2025 periastron passage will be of great interest.

The long-term light curve shows hints of variability over the orbit with a suppression of flux away from periastron and a small rise towards it, but is still compatible with a constant flux. 
On shorter time scales during the periastron passage itself, no significant variability could be observed.

Using simultaneous \fermi\ data a multi-wavelength SED was derived for the 2020 periastron passage. The combined spectrum was compared to the multi-component model by \citet{White2020}. With modifications of the position of the emission region and the cut-off energy of the hadronic component, the model matches the observed flux points well. The \hess\ data suggest, that the emission originates further out in the system at distances of ${\sim} 10$ to ${\sim} 20$~au and that the proton spectrum has a higher cut-off energy at ${\sim} 1.5$~TeV. The good match of the resulting model is in line with the predominantly hadronic origin of the gamma-ray emission from \etacar, with the \hess\ detection tracing the protons accelerated on the side of \etacar-B. 

By using such models and estimating the maximum energy with the Hillas limit, the detected signal from TeV cosmic rays provides clear evidence for the existence of a fast wind in the system consistent with the previous interpretation of X-ray data \citep[e.g.][]{EtaCar:Pittard02,Parkin_et_al_2011}.

Future measurements of the VHE gamma-ray emission during periastron passages by the Cherenkov Telescope Array Observatory (CTAO), employing stereo analysis techniques for the full energy range, could provide further insight into the wind conditions and shock physics of this unique binary system. We anticipate that \hess\ could already constrain the orbit-to-orbit VHE variability with observations around the 2025 periastron passage.

\begin{acknowledgements}
The support of the Namibian authorities and of the University of
Namibia in facilitating the construction and operation of H.E.S.S.
is gratefully acknowledged, as is the support by the German
Ministry for Education and Research (BMBF), the Max Planck Society,
the Helmholtz Association, the French Ministry of
Higher Education, Research and Innovation, the Centre National de
la Recherche Scientifique (CNRS/IN2P3 and CNRS/INSU), the
Commissariat à l’énergie atomique et aux énergies alternatives
(CEA), the U.K. Science and Technology Facilities Council (STFC),
the Irish Research Council (IRC) and the Science Foundation Ireland
(SFI), the Polish Ministry of Education and Science, agreement no.
2021/WK/06, the South African Department of Science and Innovation and
National Research Foundation, the University of Namibia, the National
Commission on Research, Science \& Technology of Namibia (NCRST),
the Austrian Federal Ministry of Education, Science and Research
and the Austrian Science Fund (FWF), the Australian Research
Council (ARC), the Japan Society for the Promotion of Science, the
University of Amsterdam and the Science Committee of Armenia grant
21AG-1C085. We appreciate the excellent work of the technical
support staff in Berlin, Zeuthen, Heidelberg, Palaiseau, Paris,
Saclay, Tübingen and in Namibia in the construction and operation
of the equipment. This work benefited from services provided by the
H.E.S.S. Virtual Organisation, supported by the national resource
providers of the EGI Federation.
\end{acknowledgements}
\bibliographystyle{aa}
\bibliography{references_cleaned}

\begin{appendix}

\section{\hess\ mono analysis - Cut on shower displacement variable}
\label{app:delta}

\begin{figure}[h]
\includegraphics[width = 0.24\textwidth]{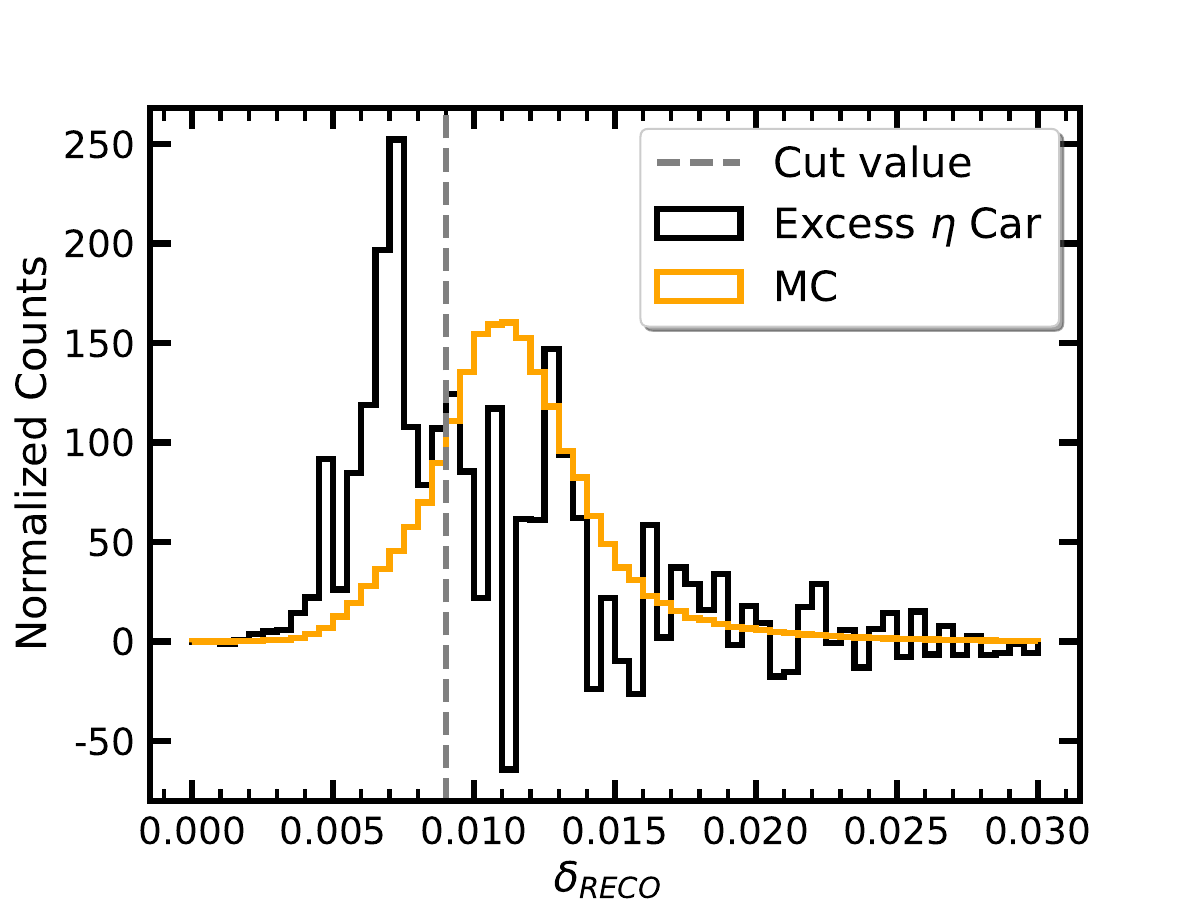}
\includegraphics[width = 0.24\textwidth]{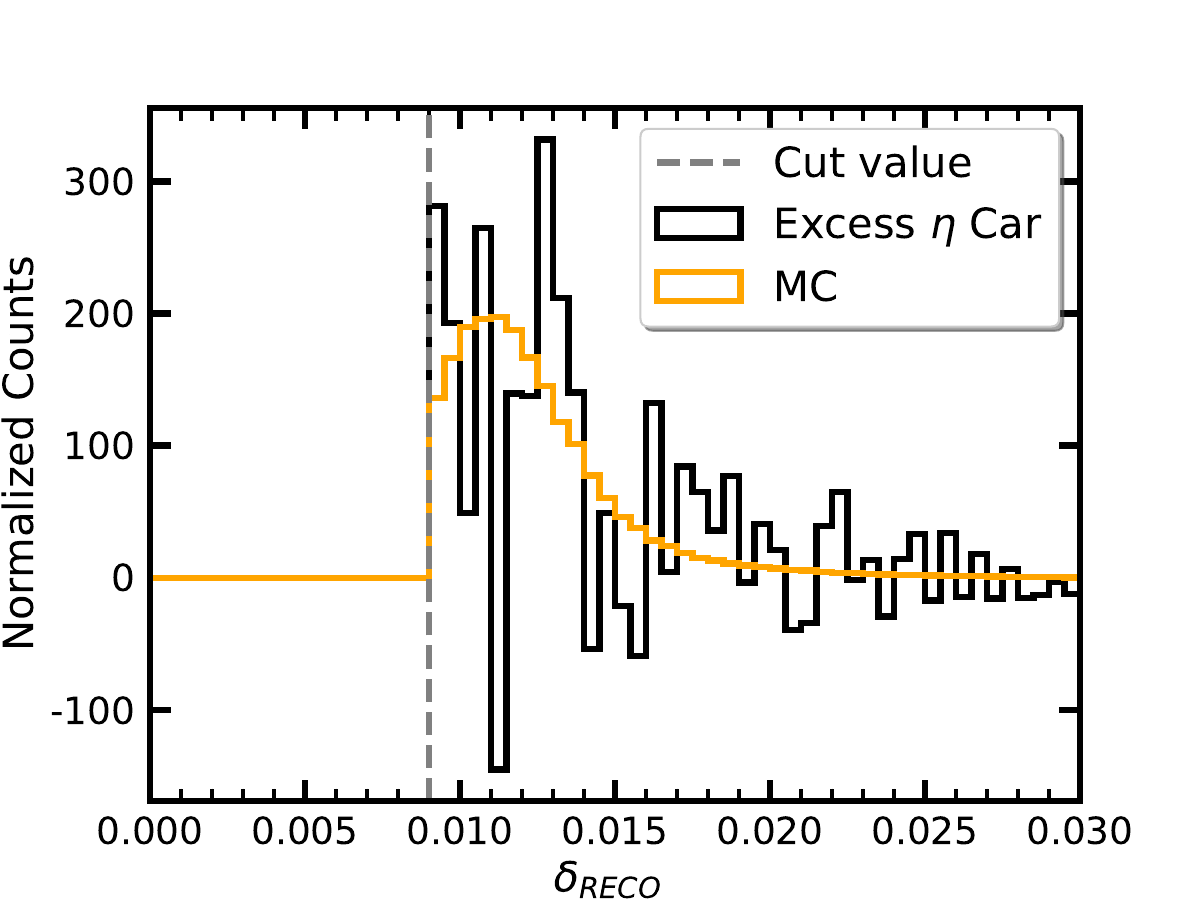}
\caption{
On the left the distributions of the displacement variable $\delta_{RECO}$ are compared between gamma simulations and excess events from the on region. The gamma simulations are point-like simulations at 40\grad\ zenith and 0.7\grad\ offset. A spectral index of 3.5 is assumed. After applying a cut of $\delta_{RECO} \geq 0.009 $ the distributions, as well as $\alpha$ have been derived again and are compared in on the right. Both the MC and excess distribution are normalised to an area under the curve of 1.
}
\label{fig:eta_deltareco}
\end{figure}

The direction of shower images influenced by high noise levels and broken pixels is hard to reconstruct with a single image. The resulting displacement parameter $\delta_{RECO}$ \citep[see][]{Murach2015} might thus be poorly constrained. Especially in the high NSB region around \etacar\ (see \autoref{fig:eta_nsbmap}), the shower direction might be reconstructed rather close to the image itself, having a small $\delta_{RECO}$. This would contribute to an excess in the region not originating from a true gamma-ray source. The $\delta_{RECO}$ distribution was derived from point source gamma simulations with 0.7\grad\ wobble offset, which were weighted to represent a source with a spectral index of 3.5. This was compared to the excess in the \etacar\ on region (radius = 0.13\grad ). The excess in the on region was derived by subtracting the normalised distribution found in the transformed off runs from the distribution found in the \etacar\ runs. The normalisation $\alpha$ was derived to have equal counts in the on and off maps outside of the exclusion mask. As presented in \autoref{fig:eta_deltareco} the distribution from gamma simulations is well constrained, whereas the excess events exhibit some excess at smaller $\delta_{RECO}$ values. Consequently, a cut on $\delta_{RECO} \geq 0.009 $ was introduced, which keeps 80 \% of the gammas but limits the number of closely reconstructed background events. The resulting distribution after applying this cut to the full DS-C data set and the corresponding off data set yields a better match. The loss in gamma efficiency was incorporated in the instrument response functions accordingly.

\section{\hess\ mono analysis comparison details}
\label{sec:app-h}

\begin{figure}
\centering
\includegraphics[width = 0.43 \textwidth]{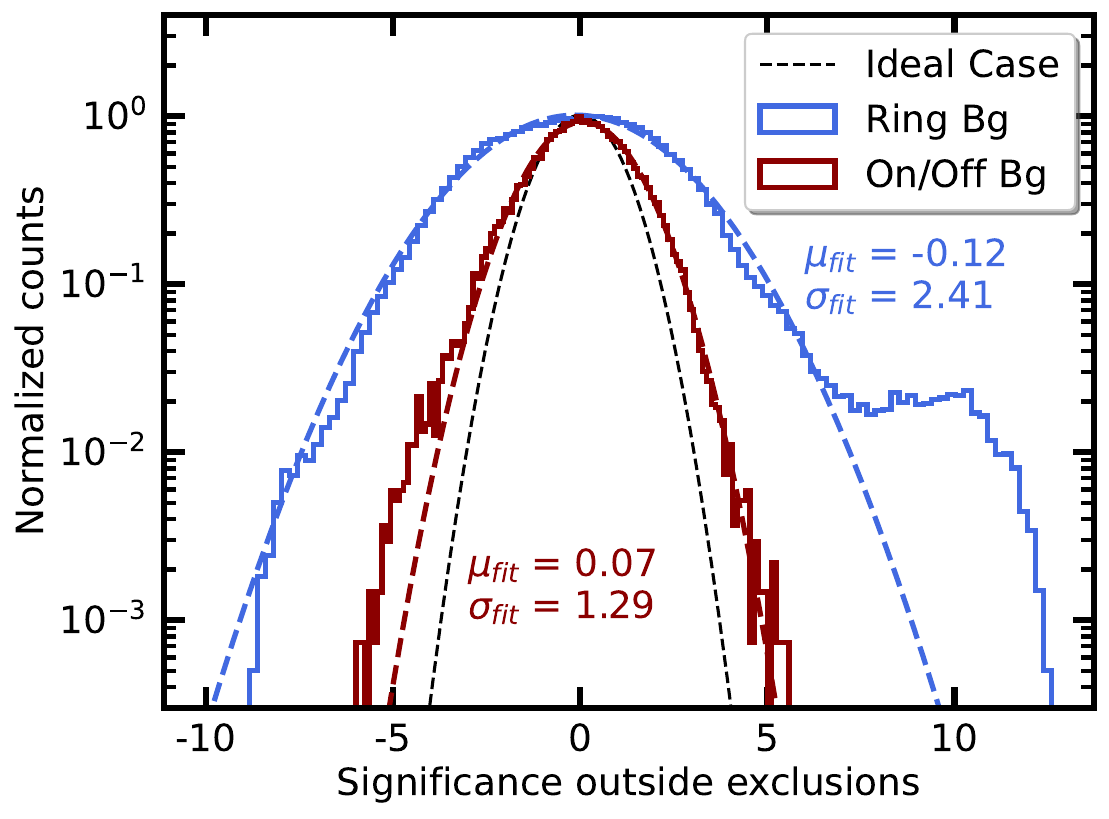}
\includegraphics[width = 0.44\textwidth]{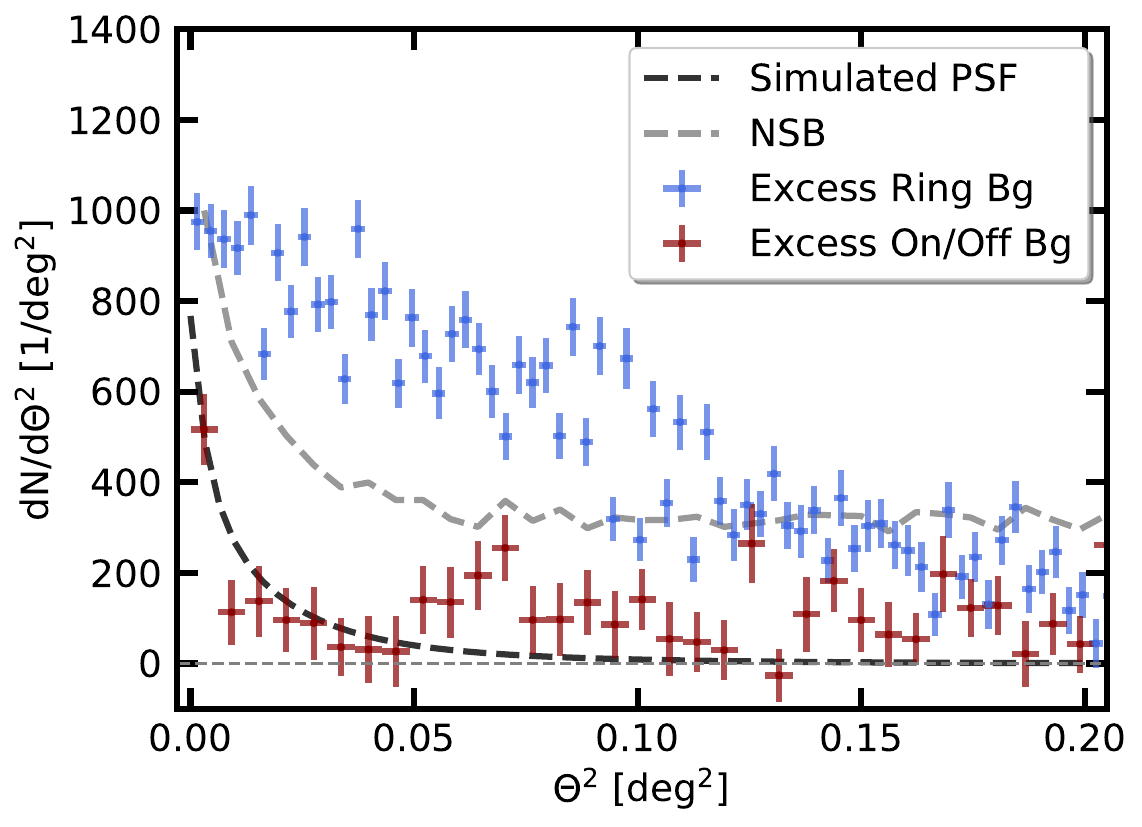}
\caption[Results of ring background mono analysis for \etacar\ ]{
\emph{Upper panel:} Resulting histogram of significance values outside of the exclusion mask (0.8\grad\ centred on \etacar )  from the ring background analysis method as also used in \citet{hessetacar} drawn together with a fitted Gaussian. The Gaussian strongly deviates from an ideal background description with width 1. For comparison the distribution of significances for the On-Off method is shown as well. \\ \emph{Lower panel:} Radial excess distribution from the ring background and the On-Off background analysis centred at \etacar\ compared with the point spread function (PSF) derived from simulations (scaled to match the On-Off excess in the first bin). Additionally, the radial NSB profile is shown with arbitrary scaling.} 
\label{fig:eta_uglymaps}
\end{figure}

For the mono analysis described in \autoref{sec:hess_mono} an On-Off background method built from Off-runs emulating the \etacar\ observing conditions was utilised. If a ring or reflected regions background technique \citep{Berge2007} is applied as it was done in \citet{hessetacar}, due to the inhomogeneous NSB, the resulting signal is contaminated by noise artefacts depicting a very large extended excess strongly deviating from the size of the point-spread function. This is not in line with the expectation of a point-like signal, as the physical scales relevant for the VHE emission \citep[see e.g.][]{Steinmassl23} are well below the resolution limit of \hess\  Furthermore, the background normalisation is very poor with obvious features in the significance map outside the exclusion regions up to $\pm 12\,\sigma$. A Gaussian fit to the distribution of these significance values yielded a width of 2.4 deviating strongly from a good background description with width 1. These findings are summarised in \autoref{fig:eta_uglymaps}. 

 The effects of noise and broken pixels, that will e.g. crop or enlarge images ultimately leading to a bad direction reconstruction, are not negligible and not homogeneously distributed. From this, it is obvious that the assumption of a radially symmetric acceptance does not hold for this particular FoV. The NSB peaks close to the position of \etacar\ with an extended high NSB region coincident with the extent of the Carina Nebula (see \autoref{fig:eta_nsbmap} and \autoref{fig:eta_uglymaps}). This strengthens the point that the excess derived with the ring background method is indeed mostly a noise artefact.

\begin{figure}[h]
\centering
\includegraphics[width = 0.45\textwidth]{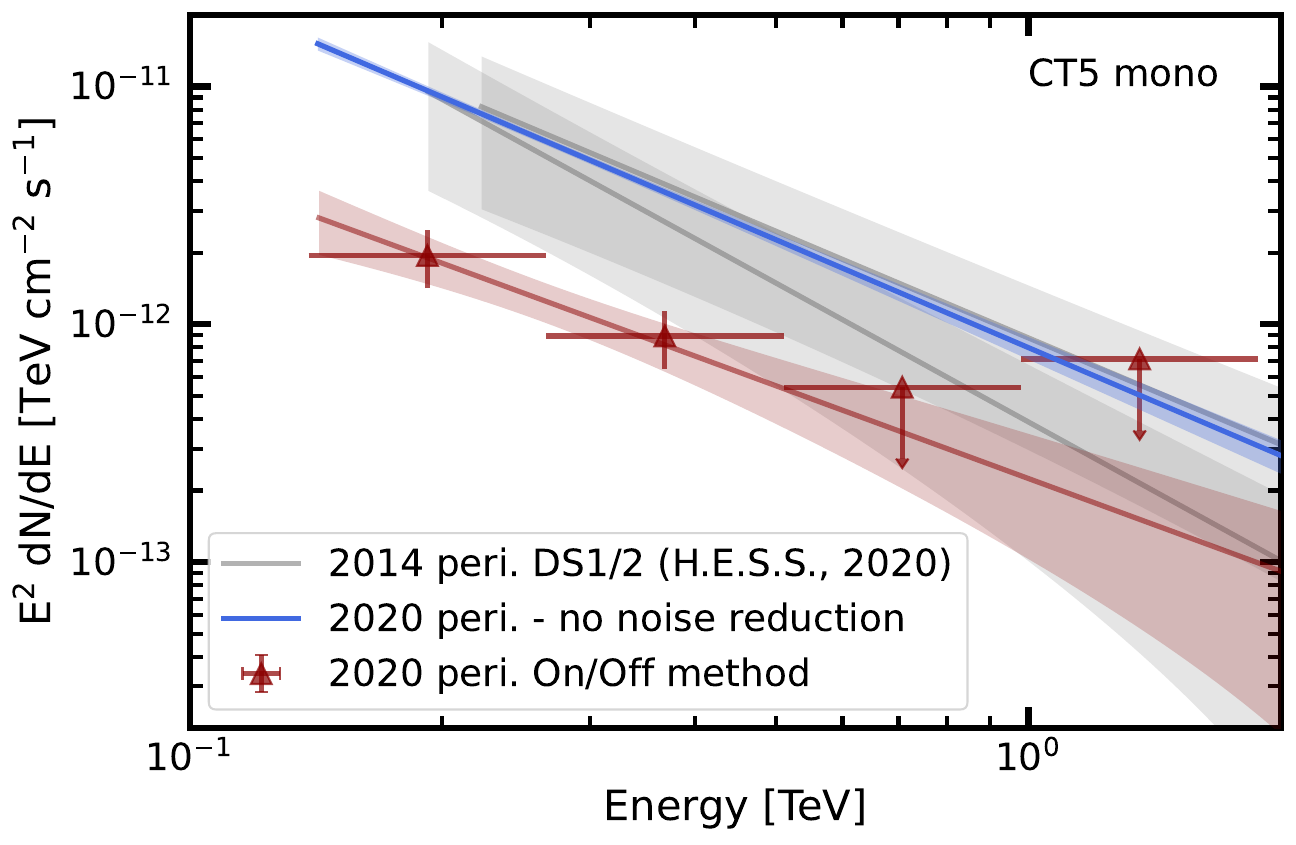}
\caption[Spectral results of the \etacar\ mono analysis in comparison]{
 The resulting SED from the On-Off approach with flux points is compared to the spectral model derived from a reflected regions analysis without any special treatment to model the noise influence. The spectra from the datasets (DS) 1 and 2 taken around the 2014 periastron passage from \citet{hessetacar} are also shown for comparison. The spectral models for the 2020 dataset (DS-C) only include statistical errors, whereas the 2014 spectra also include systematic errors as discussed in the paper.}
\label{fig:eta_spectralderivation}
\end{figure}

A spectrum derived with the default reflected regions background method resulting in a similar large signal as with the ring background method yields a flux estimate considerably higher than the one derived with the careful On-Off background treatment. The two flux levels are inconsistent but the result of the reflected region analysis is consistent with the results published in \cite{hessetacar} (see \autoref{fig:eta_spectralderivation}). This strongly implies that the spectrum presented there suffers from unaccounted systematics due to noise.

\section{Further \fermi\ analysis details}
\label{sec:app-a}

\begin{table}[H]
\caption{Configuration used for the \fermi\ analysis}.
    \centering
    \begin{tabular}{l|l}
    	
    	Parameter &Value \\
    	\hline \\
    	Data release & P8R3 \\
    	IRFs & P8R3\_SOURCE\_V3 \\
    	ROI data width & 20\grad \\
    	ROI model width & 25\grad \\
    	Bin size & 0.1\grad\ \\
    	zmax & 90\grad\ \\
    	Coordinate system & GAL \\
    	Minimum energy & 100 MeV \\
    	Maximum Energy & 500 GeV \\
    	MET start & 239557417 /  598755697 \\
    	MET stop & 688521600 / 612045512 \\
    	MET excluded (ASASSN-18fv) & 542144904 – 550885992 \\
    	evclass & 128 \\
    	evtype & 3 \\
        edisp & True \\
        edisp binning & -1 \\
        edisp disabled & Isotropic \\
        Light curve binsize & 5244652.8 s\\
    	Galactic diffuse template & gll\_iem\_v07.fits \\
    	Isotropic background component & \scriptsize{iso\_P8R3\_SOURCE\_V3\_v1.txt} \\ 
    	\fermi\ catalogue & 4FGL-DR3 (\small{gll\_psc\_v29.fits}) \\
    \end{tabular}
   \tablefoot{The time period MET 542144904 – 550885992 was excluded due to the bright nova ASASSN-18fv, or V906 Carinae, \citep{Aydi20} within the ROI.}
    
    \label{tab:fermi_cfg}
    \end{table}
\newpage 
    
\section{Further details on the model}
\label{sec:app-model}
As discussed in \autoref{sec:int} the predicted gamma-ray spectra before absorption were taken from \citet{White2020}.
In their model, diffusive shock acceleration at the two shocks associated with the two winds is calculated for several phases based on the model described in \citet{Ohm2015}. The model adopts a highly eccentric orbit of the system (see \autoref{tab:orbit}) and the stellar parameters as summarised in \autoref{tab:whitemodel}.

The individual winds are assumed to be constant, spherically symmetric and to have reached terminal velocity before the wind collision region. The geometry of the shock cap itself is governed by the momentum balance of the two winds. The parameters used to model the efficient acceleration in the system are summarised in \autoref{tab:whitemodel}. It is assumed that ${\sim} 10$ \% of the available wind power goes into the acceleration of protons on each shock.

In all cases, the electron acceleration is limited by inverse Compton losses. Furthermore, all non-thermal particles accelerated at the shock of \etacar-A are calorimetric, due to the small energy-loss timescale (${\sim}$10 days) for interactions of accelerated protons with nuclei in the dense wind. 
The protons accelerated on the more tenuous wind of \etacar-B eventually escape the WCR ballistically and can interact where the mixing of the two winds occurs. As discussed in more detail in \citet{Steinmassl23}, it is estimated that approximately $90 $\% of cosmic rays accelerated at the shock of \etacar-B eventually escape, resulting in a power of escaping cosmic rays of $6.5 \times 10^{35} \UNITS{erg\,s^{-1}}$.

\citet{White2020} assumed in their paper a mixing length of one shock cap radius. This implies a small emission region during periastron and was used in their paper to calculate the absorption behaviour due to gamma-gamma interactions with the stellar photon fields. The \hess\ measurements constrain the absorption and hence the size of the emission region, which was further investigated in this work.

\begin{figure}[]
\centering
\includegraphics[width = 0.45\textwidth]{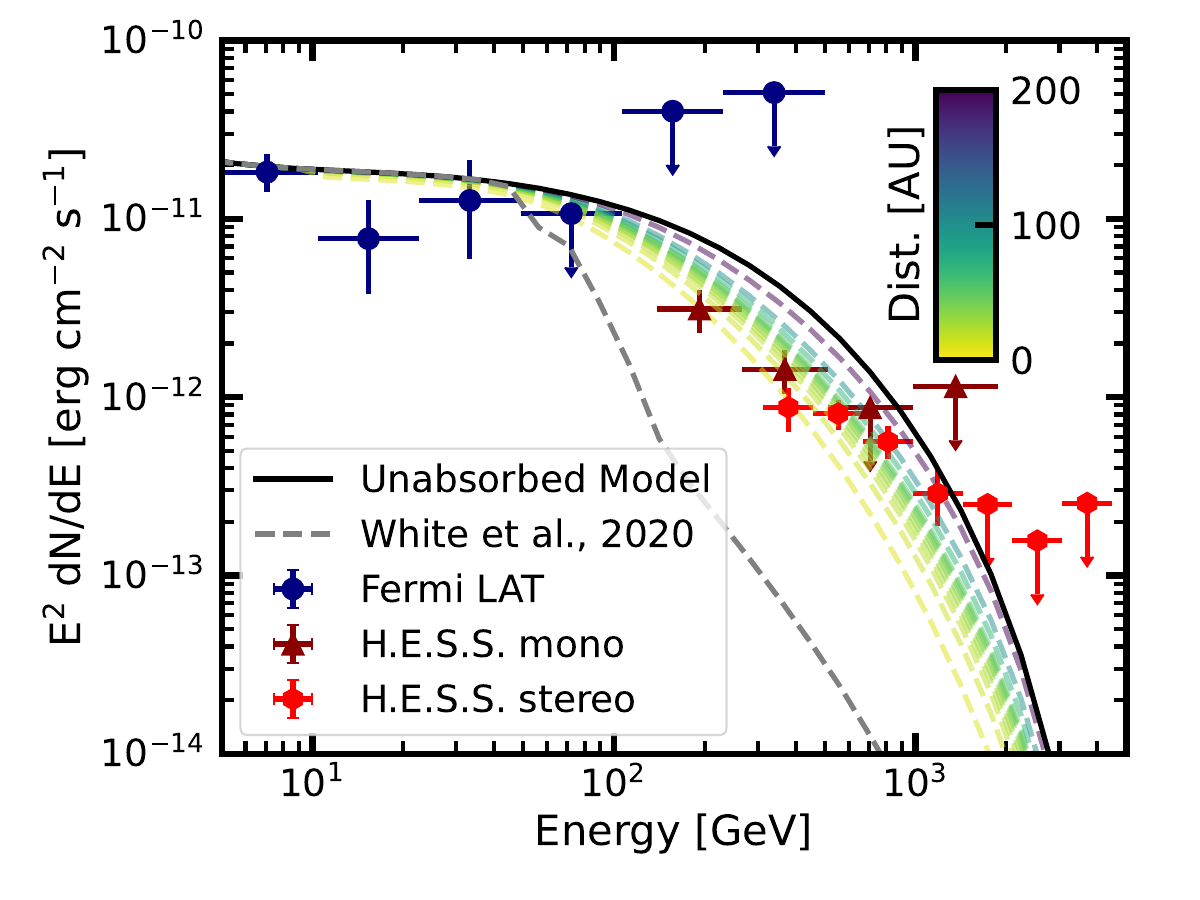}
\includegraphics[width = 0.45\textwidth]{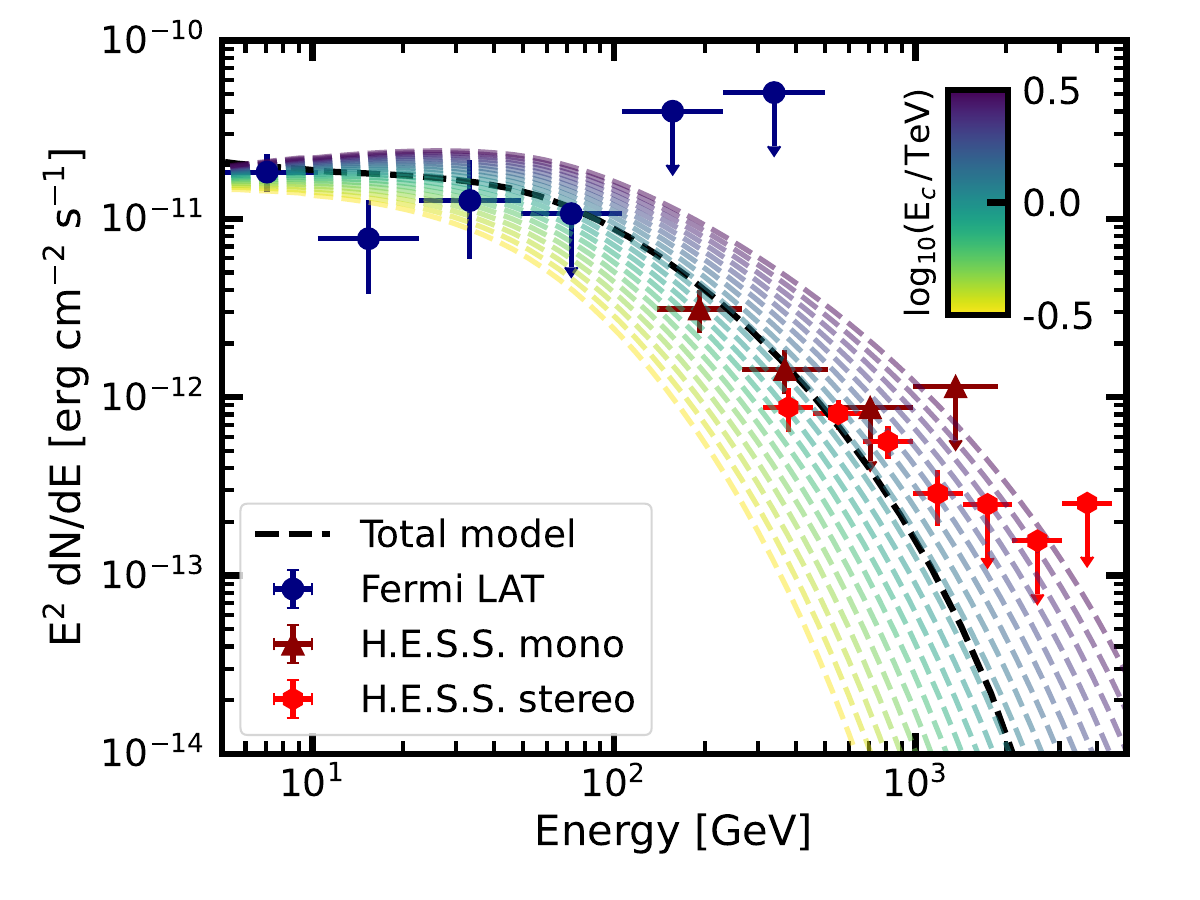}
\caption{Model spectra for different emission regions and cut-off energies compared to the combined SED. \\
\emph{Upper panel:} The total predicted model flux for the periastron phase is compared to the data assuming different gamma-gamma absorption curves. For comparison, also the absorbed and unabsorbed model from \citet{White2020} (as also presented in \autoref{fig:eta_model_20au}) is shown. The absorption in their model assumes gamma-ray production close to the shock cap. \\ \emph{Lower panel:} The cut-off energy $E_c$ is varied, keeping the index and normalisation fixed. The cut-off energy was varied in 20 logarithmic bins between 0.32 and 3.2 TeV. The emission region was assumed to be at a distance of 20~au and the resulting absorption curve was used. The total model refers to the total emission from the original model without adapting $E_c$.}
\label{fig:eta_modelabs}
\end{figure}

Therefore, we utilised the predicted gamma-ray spectra before absorption during the periastron phase (phase 0.995 to 1.025 in their work). They were then matched with different transparency curves as a function of energy with the same method as in \citet{Steinmassl23}. 
The characteristics of the photon fields are summarised as well in \autoref{tab:whitemodel}. For \etacar-A two different variants were assumed, one following the higher temperature value used in \citet{White2020} and another one using the lower temperature derived by \citet{Groh2012}. The transparency curves were calculated for spheres at radial distances of 10 to 200 au. As a result, the total gamma-ray emission assuming the different absorption curves is presented in \autoref{fig:eta_modelabs}. There, the results obtained with the lower temperature case for \etacar-A are presented. Nevertheless, even though the shape of the transparency curve changes between the two variants, in both cases an emission region at distances of $\sim$10 to $\sim$20~au is needed to explain the VHE flux level. 

To account for a slightly higher flux in the low-energy component in comparison with the dataset in \citet{White2020} based on the previous periastron passage, the flux from \etacar-A was increased by a factor of 1.1. 

Furthermore, the cut-off energy of the proton spectrum with index 1.8 was estimated to be 1.1 TeV in the model of \citet{White2020}. This estimate was obtained by fitting an exponential cut-off power law proton spectrum of the form $\phi = \phi_0 \left( \frac{E}{E_0} \right) ^{-\Gamma} \exp{\left( -\left( \frac{E}{E_c} \right)^{\alpha} \right)}$ using the Gamera package \citep{gamera_2015,gamera_2022}. The normalisation $\phi_0$ was fixed to be equal to the model spectrum at 1 GeV, the cut-off index set to  $\alpha =1$ and the fit was run between 1 GeV and 3 TeV. As this cut-off can be better constrained by the \hess\ measurements presented in this paper the proton cut-off energy was varied in 20 logarithmic bins between 0.32 and 3.2 TeV keeping all other parameters fixed. The corresponding gamma-ray spectra were calculated using again the Gamera package and compared to the periastron SED (see also \autoref{fig:eta_modelabs}) applying the absorption curves at 20 au. This implies a cut-off energy for protons of ${\sim}$1.5\,TeV to explain the shape of the VHE data.

\begin{table}[]
\caption{Assumed orbital parameters}
    \centering
    \begin{tabular}{l | c c }
    Parameter   & Value  & Reference \\ 
    \hline \\
    Eccentricity $\epsilon$ & 0.9 & 1 \\
    Semi-major axis a [au] & 16.64 & 2 \\
    Inclination [\grad ] & 135 & 3 \\
    Position angle [\grad ] & 10 & 3 \\
    \hline
    \end{tabular}
    \tablebib{(1)~\citet{Damineli2008}; (2) \citet{Hillier2001}; (3) \citet{Madura2012}.}
    
    \label{tab:orbit}
    \end{table}
    
\begin{table}[]
\caption{Stellar and model parameters}
    \centering
    \begin{tabular}{l | c c c}
    Parameter   & \etacar -A    & \etacar -B  & References \\ 
    \hline \\
    $R_{\star}$ [\si{\solarradius}]     & \num{100}  \slash \, \num{843}  & \num{20}    & 1,2 \\
    $T_{\star}$ [\SI{e4}{\kelvin}]    &  \num{2.58} \slash \, \num{0.94}   & \num{3.0}     & 3,2  \\
    $L_{\star}$ [\SI{e6}{\solarluminosity}] & \num{4} \slash \, \num{5}  & \num{0.3} & 3,2 \\
    $\dot M$ [\si{\solarmass\per\year}] & \num{4.8e-4}  & \num{1.4e-5} & 4\\
    $v_{\infty}$ [\si{\kilo\meter\per\second}]  & \num{5e2} & \num{3e3} & 4\\  
    $B_{\star}$ [\si{\gauss}]  & \num{100} & \num{100} & \\
    $v_{\text{rot}}$ [$v_{\infty}$] & \num{0.15}    & \num{0.15} & \\
    $\eta_{\rm acc}$          & \num{15}  &   \num{5} & \\
    $P_\text{p}$ [$P_\text{wind}$]    & \SI{10}{\percent}  & \SI{9}{\percent} & \\
    $P_\text{e}/P_\text{p}$ &  \SI{3}{\percent}   & \SI{3}{\percent}  & \\
    \hline
    \end{tabular}
   \tablefoot{Adapted from Table 1 in \citet{White2020}. $R_{\star}$ denotes the stellar radius, $T_{\star}$ the surface temperature, $L_{\star}$ the luminosity, $\dot M$ the mass loss rate and $v_{\infty}$ the thermal velocity. The surface magnetic fields $B_{\star}$, the rotation velocities $v_{\text{rot}}$, the acceleration efficiency parameters $\eta_{\rm acc}$, the wind power going into the acceleration of protons  $P_\text{p}$ and the ratio $P_\text{e}/P_\text{p}$ of the power going into the acceleration of electrons above \SI{1}{\mega\electronvolt} and protons above \SI{1}{\giga\electronvolt} were chosen in their model. For the photon field of \etacar-A two different scenarios were assumed.}
   \tablebib{(1)~\citet{Corcoran_Hamaguchi2007}; (2) \citet{Groh2012}; (3) \citet{Davidson1997}; (4) \citet{EtaCar:Parkin09}.} 
    
    \label{tab:whitemodel}
    \end{table}
    
\end{appendix}
    
\end{document}